\documentclass[sn-mathphys-num]{sn-jnl}

\usepackage{graphicx}
\usepackage{multirow}
\usepackage{amsmath,amssymb,amsfonts}
\usepackage{booktabs}
\usepackage{url}

\providecommand{\arcsec}{\ensuremath{^{\prime\prime}}}
\begin{document}

\title[CNN search for USPs and blends among TESS false positives]{%
A CNN search for ultra-short-period planets in TESS: blends among real
false positives and an independent validation of TOI-6662~b}

\author*[1]{\fnm{Juan Francisco} \sur{Pag\'es Navarrete}}
\email{jfpages@gmail.com}
\affil*[1]{\orgname{Independent researcher}, \city{M\'alaga},
           \country{Spain}}

\abstract{%
We trained a simple convolutional neural network, using a phase-folded
global view only, to detect ultra-short-period planets in TESS photometry,
initially on synthetic trapezoidal transits and synthetic eclipsing
binaries. Evaluated against catalogued eclipsing binaries it reaches an area
under the ROC curve of 0.97. Keeping the same model and the same planets, but
replacing the negatives with false positives rejected by the TESS Follow-up
Observing Program, the area falls to 0.50. Benchmarks built from catalogued
or simplistic synthetic binaries can therefore greatly overstate operational
performance, a point long recognised in the field, which is why the leading
classifiers are trained and tested on real vetted false positives.

Of 145 such false positives examined at pixel level, at least 42\% are
blends with a contaminating source more than one TESS pixel away. Under the
simplifying assumption that these are indistinguishable from transits in the
light curve, this fraction implies a heuristic bound of $1-f/2 \approx 0.79$
on the AUC of a light-curve-only classifier; fine-tuning our network on real
false positives reaches $0.80 \pm 0.06$, consistent with, but not a strong
test of, that bound. These figures characterise our detector and should not
be read as a limit on classifiers using richer inputs.

A detector alone is therefore insufficient. We built CASCADE-USP, a
pipeline chaining the network with difference-image centroid analysis,
scene modelling with high-resolution imaging and the Gaia astrometric
solution, supplying in sequence what the light curve lacks. Its output is a
prioritisation, not a verdict: it records what evidence each candidate
still needs. Only the classifier is domain-specific; the other stages apply
to any transit candidate. Of 435 unresolved TESS candidates it isolates 20;
19 prove false positives, traced to causes invisible in TESS photometry.
The twentieth is TOI-6662~b, an Earth-sized planet
($0.94 \pm 0.03\,R_\oplus$) orbiting an M dwarf every 7.05 hours. Our
validation, independent of a concurrent one, agrees with it in planetary parameters and classification, closing the CASCADE
procedure with an external check.

\keywords{planets and satellites: detection,
          planets and satellites: terrestrial planets,
          methods: statistical,
          techniques: photometric,
          stars: individual: TOI-6662}

}
\maketitle

\section{Introduction}\label{sec:intro}

\subsection{Ultra-short-period planets}\label{sec:intro-usp}

The study of ultra-short-period planets opens with CoRoT-7b
\citep{Leger2009}, the first transiting super-Earth with a measured radius,
on a 0.854~d orbit around a G9 dwarf, and with Kepler-10b
\citep{Batalha2011}, the first rocky planet identified by \textit{Kepler}.
Both established the existence of a population of terrestrial planets on
orbits of hours, subject to conditions without analogue in the Solar
System.

Ultra-short-period planets (USPs), conventionally defined by an orbital
period below one day \citep{Winn2018}, constitute a singular population.
They orbit at distances of a few stellar radii, where they receive
irradiations of hundreds to thousands of times the terrestrial value and
experience temperatures sufficient to melt silicates on their dayside. They
are, almost without exception, of terrestrial size or smaller: the radius
distribution falls sharply above $\sim$2~$R_\oplus$, a consequence of any
gaseous envelope being stripped by photoevaporation in this irradiation
regime \citep{Lundkvist2016,Fulton2017}. Their proximity to the host
further places them close to the tidal disruption limit, which imposes
lower bounds on their density \citep{Rappaport2013}.

Their interest is twofold. They offer direct access to the composition of
planetary cores stripped of atmosphere --- GJ~367~b, with
$0.72\,R_\oplus$ and an iron fraction comparable to that of Mercury, is the
paradigmatic example \citep{Lam2021} --- and their high transit frequency
makes them favourable targets for characterisation. Their occurrence rate
peaks around low-mass stars \citep{SanchisOjeda2014}, so that M dwarfs
constitute the natural hunting ground. The population remains small: only
around a dozen ultra-short-period rocky planets orbiting M dwarfs have both
mass and radius determined \citep{Srivastava2026}.

\subsection{Deep learning for transit detection}\label{sec:intro-dl}

Identifying transits in photometric surveys is an imbalanced classification
problem over data volumes that make exhaustive inspection unfeasible. Since
the foundational work of \citet{Shallue2018}, convolutional networks have
become the reference tool, with successive developments incorporating
domain knowledge \citep{Ansdell2018}, adapting to TESS data
\citep{Yu2019,Osborn2020,Tey2023}, or, together with other
machine-learning classifiers, attaining statistical validation
capability \citep{Armstrong2021,Valizadegan2022,Valizadegan2023}. In the
ultra-short-period regime, the GPFC method combines hardware-accelerated
phase folding with convolutional classification \citep{Wang2024}. Recent
developments have diversified both architectures and inputs, with
transformer-based models applied to full-frame images \citep{Salinas2025}
and classifiers incorporating background flux and centroid information
\citep{TardugnoPoleo2024}.

The declared performance of these systems is high: AUC values of
0.988 \citep{Shallue2018}, 0.999 \citep{Armstrong2021} and 1.0
\citep{Valizadegan2022} have been reported. These figures were obtained on
real \textit{Kepler} threshold-crossing events with dispositions from
vetting and follow-up, so that their negative classes include genuine
astrophysical false positives such as background eclipsing binaries; several
of these classifiers further use centroid and stellar information as
inputs \citep{Ansdell2018,Valizadegan2022}. Where simulated training sets are
used, the most successful approaches rely on detailed simulations that include
blended scenarios, either at pixel level \citep{Osborn2020} or injected into
real TESS light curves, as in the RAVEN pipeline \citep{Lafarga2026}.

\subsection{The benchmarking problem}\label{sec:intro-bench}

Not every classifier, however, is developed under those conditions.
Detectors trained on simple synthetic light curves, or evaluated against
catalogued eclipsing binaries, are still common outside the main pipelines,
and for them the evaluation set may describe a substantially easier problem
than the operational one.

The concern is not new. \citet{Osborn2020} warned explicitly that
transferring a model trained on simulations to real data would not perform
with equal accuracy, and the review of \citet{HuertasCompanyLanusse2023}
identifies the shift from the simulated to the observational domain as a
potential source of catastrophic failure. In the machine-learning
literature the phenomenon is characterised: \citet{Geirhos2020} term it
\textit{shortcut learning}, decision rules that perform well on standard
benchmarks but fail to transfer to more demanding conditions, a
manifestation of the ``Clever Hans'' effect described by
\citet{Lapuschkin2019}. Systematic biases introduced by deep networks when
annotating large astronomical catalogues have been described in general
terms by \citet{DharShamir2022}. Our aim here is the more modest one of quantifying that
degradation for a concrete detector in the ultra-short-period regime, and of
measuring how much of it is attributable to blends.

A major contributor is the \textit{blend}: genuine eclipses
originating in a star other than the target and diluted within the
photometric aperture. With a pixel scale of 21~arcsec \citep{Ricker2015},
TESS apertures routinely contain flux from several sources, so that the
contaminating signal can closely mimic a shallow transit in the flux time
series. Much of the information that allows discrimination --- the colour,
the position of the source, the surrounding stellar field --- does not
reside in the light curve. \citet{Twicken2018}, describing the
difference-image diagnostic of the Data Validation module, state the point
directly: the technique exploits spatial information contained in the pixel
data, information that is not available in photometric flux time series.

\subsection{This work}\label{sec:intro-this}

We present the construction, characterisation and application of a
simple convolutional detector for ultra-short-period planets in TESS, organised
around three contributions.

\paragraph{First, methodological.}
We quantify, for our detector, how much evaluation against catalogued
eclipsing binaries overstates performance on real false positives. Through a factorial design that varies the positive and
negative classes separately, we show that the AUC of our detector falls
from 0.97 against catalogued binaries to 0.50 --- the chance value ---
against the real false positives of the TESS Follow-up Observing Program,
and that the degradation is attributable mainly to the nature of the
negatives.

\paragraph{Second, physical.}
We measure the contribution of blends. Difference-image analysis of 145
real false positives shows that at least 42\% are blends with a source more
than one TESS pixel away. A heuristic bound derived from that fraction is
consistent with the AUC attained after retraining with real negatives,
although the comparison is not a strong test (Sect.~\ref{sec:ceiling}).

\paragraph{Third, observational and practical.}
We formalise the resulting procedure in \textbf{CASCADE-USP}
(CNN-Assisted Sequential Centroid And Dilution Evaluation), a pipeline
chaining photometric classification with difference-image centroid analysis
and dilution evaluation through TRICERATOPS \citep{Giacalone2021} and
high-angular-resolution contrast curves, the latter located and retrieved
automatically from ExoFOP. Applied to 435 unresolved TOI candidates,
CASCADE-USP recovers 20 objects after spatial filtering, of which we
document 19 false positives individually by failure mode and statistically
validate one: \textbf{TOI-6662~b} (TOI-6662.01), an Earth-sized planet
($0.94 \pm 0.03\,R_\oplus$) on a 7.05-hour orbit around an M dwarf.
The same object has been validated by \citet{Collier2026} in an
automated TRICERATOPS analysis of 443 TESS candidates. The two analyses
are independent\footnote{This manuscript was first submitted to
\textit{Experimental Astronomy} on 13 August 2026, before the public
release of \citet{Collier2026} (arXiv:2608.21560, 21 August 2026).} and
reach the object by different routes; their agreement
(Sect.~\ref{sec:collier}) is the last step of the CASCADE procedure: an
external check of the full pipeline on a real planet.

The design of CASCADE responds directly to the diagnosis of this work.
Since a substantial fraction of false positives cannot be reliably
resolved from the light curve, the pipeline does not pursue a better classifier but the
systematic incorporation of the information channels that photometry does
not contain. Each stage contributes one of them: source position through
pixel-level analysis \citep{Bryson2013,Twicken2018}, and the stellar field
and unresolved companions through scenario modelling with TRICERATOPS
\citep{Giacalone2021} and speckle imaging \citep{Scott2021}. We further
add a fourth channel, independent of the preceding three: the Gaia
DR3 astrometric solution quality \citep{Lindegren2021}, which flagged one
false positive that no other channel detected. The output is not a verdict
but a decision table which, for each candidate, additionally records
\emph{what information is missing for validation}, allowing observing time
to be prioritised.

The code, datasets and intermediate results supporting this article are
published to permit reproduction and reuse. The framework is independent of
the period regime: only the classification stage is domain-specific, so
that adaptation to other populations reduces to retraining the classifier.

The article is organised as follows. Sect.~\ref{sec:detector} describes the
detector and documents a spurious shortcut learned during its development,
which prefigures the central result. Sect.~\ref{sec:collapse} quantifies
the performance collapse against real false positives and diagnoses its
cause. Sect.~\ref{sec:retrain} presents the retraining and its comparison
with a heuristic bound derived from the blend fraction. Sect.~\ref{sec:application} describes the
application to unresolved candidates and the false-positive taxonomy.
Sect.~\ref{sec:planet} presents the validation and characterisation of
TOI-6662~b and compares them with those of \citet{Collier2026}. Sects.~\ref{sec:discussion} and \ref{sec:conclusions} discuss
the implications and summarise the conclusions.

\section{The detector}\label{sec:detector}

\subsection{Architecture and input representation}\label{sec:arch}

The classifier is a one-dimensional convolutional neural network operating
on the phase-folded light curve. The input consists of 256 phase bins,
standardised to zero mean and unit variance, with the transit minimum
cyclically shifted to the central bin.

The network comprises four successive convolutional blocks --- with 16, 32,
64 and 64 channels, and kernels of size 5, 5, 3 and 3 respectively --- each
followed by batch normalisation \citep{IoffeSzegedy2015} and ReLU
activation; the first three additionally incorporate max pooling with
factor 2. Adaptive average pooling reduces the representation to a
64-component vector, which feeds a dense classifier head of 64 units with
dropout regularisation \citep[$p = 0.3$;][]{Srivastava2014} and a sigmoid
output. The model totals of order $5 \times 10^4$ adjustable parameters.
The design follows the convolutional approach established for folded
light-curve classification \citep{Shallue2018,Yu2019}.

\subsection{Training on synthetic data}\label{sec:synth}

The training set comprises $2 \times 10^5$ synthetic curves balanced
between positives and negatives. Positives are generated as trapezoidal
profiles with variable depth, width and flat-bottom ratio, on Gaussian
noise. Negatives combine four populations: synthetic eclipsing binaries
(40\%), sub-threshold transits (20\%), sinusoidal stellar variability
(12\%) and pure noise (28\%).

The use of synthetic sets for training is widespread in this domain:
\citet{Osborn2020} employed pixel-level simulations generated with Lilith,
\citet{ZuckerGiryes2018} evaluated convolutional networks for shallow
transit detection on simulated data, and the GPFC method trains on two
million synthetic curves \citep{Wang2024}.

The generation of the synthetic eclipsing binaries proved to be the
critical element of the design, for reasons set out below.

\subsection{Generator evolution and diagnosis of a spurious
shortcut}\label{sec:shortcut}

We document here three successive iterations of the negative generator,
whose outcome constitutes the first indication of the problem developed
systematically in Sect.~\ref{sec:collapse}.

\paragraph{First version (v1).}
Negatives consisted mostly of noise and variability, without realistic
eclipsing binaries. Evaluated on a set of 513 real objects, the detector
reached an AUC of 0.80, with a recall of 52\% at the cost of a 28\%
false-positive rate: excessively permissive, it admitted binaries readily.

\paragraph{Second version (v2).}
Synthetic eclipsing binaries with pronounced V-shaped profiles (flat-bottom
ratio between 0 and 0.4) and secondary eclipses were introduced. The
synthetic AUC rose to 0.99, but behaviour on real data proved erratic: the
recall fell below 25\%, with confirmed ultra-short-period planets receiving
null scores.

\begin{figure}[t]
\centering
\includegraphics[width=\textwidth]{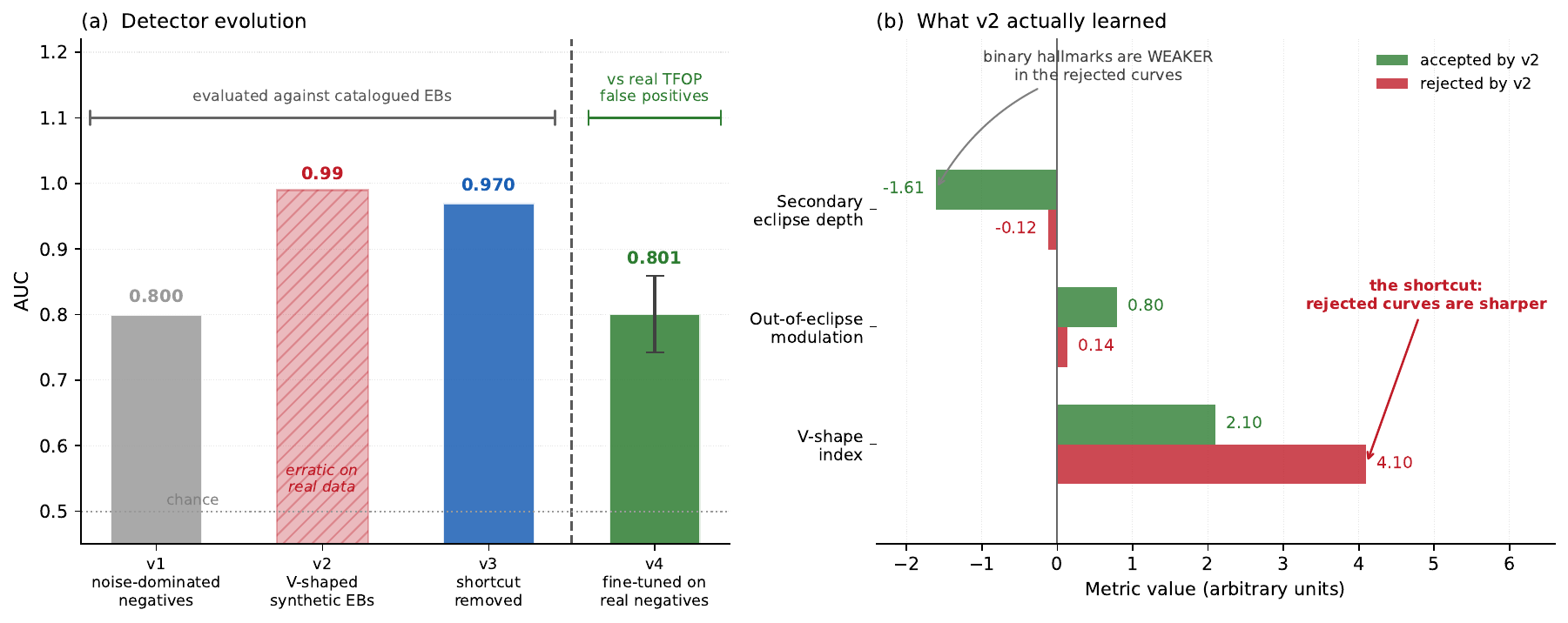}
\caption{Detector evolution and the geometric shortcut. Panel (a): AUC
across the four versions. The vertical divider marks a change of evaluation
set: v1--v3 are scored against catalogued eclipsing binaries, v4 against
real TFOP false positives. The two benchmarks are not comparable --- their
non-comparability is the subject of Sect.~\ref{sec:collapse} --- and are
separated for that reason. Version v2 attained a synthetic AUC of 0.99
while behaving erratically on real data. Panel (b): the morphological
diagnostic that revealed the cause. Curves rejected by v2 display
\emph{weaker} binary hallmarks than those accepted --- shallower secondary
eclipses and smaller out-of-eclipse modulation --- but are markedly
sharper: the classifier had learned that a narrow, peaked transit indicates
an eclipsing binary, a regularity of the generator rather than of the
physics.}
\label{fig:shortcut}
\end{figure}

The diagnosis was made by comparing the morphological properties of
rejected against accepted curves, and yielded a result contrary to the
initial hypothesis (Fig.~\ref{fig:shortcut}). The rejected objects did
\emph{not} display the physical hallmarks of a binary: their secondary
eclipses were shallower ($-0.12$ against $-1.61$ for the accepted ones) and
their out-of-eclipse modulation smaller (0.138 against 0.797). What did
distinguish them was the sharpness of the primary transit: the rejected
curves were markedly more peaked (V-shape index of 4.10 against 2.10).

The cause proved to be a generator artefact. By constructing the synthetic
binaries systematically as narrow V shapes, we had given them a
morphological marker that the classifier learned to exploit: \emph{the
network learned that ``narrow, sharp V equals eclipsing binary'' rather
than the physically discriminating signature, which is the presence of a
secondary eclipse and ellipsoidal modulation} --- the standard diagnostics
in binary classification \citep{Prsa2022}. The problem is that many real
ultra-short-period planets produce, when folded, transits that are equally
narrow and peaked, a consequence of the small number of cadences falling
within the transit, so that they were classified as binaries.

The phenomenon corresponds to what \citet{Geirhos2020} term
\textit{shortcut learning}: decision rules that perform satisfactorily on
the usual evaluation sets but fail to transfer to more demanding
conditions. It is a manifestation of the ``Clever Hans'' effect described
by \citet{Lapuschkin2019}, in which a model attains high performance by
exploiting a spurious correlation of the training set instead of the
intended causal relation. In our case the spurious correlation was
introduced inadvertently by the negative-generation procedure itself.

\paragraph{Third version (v3).}
The correction consisted in deliberately removing the shortcut, making the
synthetic binaries \emph{indistinguishable from planets in primary-transit
morphology}: the same width distribution, the same range of flat-bottom
ratio, and overlapping depths. Binaries are then distinguished solely by
the two features that physically separate them from a planet --- the
secondary eclipse at mid-phase and the ellipsoidal modulation --- with each
guaranteed to display at least one of them to avoid label noise.

Evaluated on the real set, v3 attains an AUC of 0.9699, with a recall of
93.6\% against a 4.6\% false-positive rate at the adopted decision
threshold (0.02). It dominates both earlier versions: it recovers more
planets than v1 with six times fewer false positives, and triples the
recall of v2.

\subsection{An irreducible false negative}\label{sec:fn}

One confirmed ultra-short-period planet, TOI-141~b
\citep[HD 213885~b;][]{Espinoza2020}, receives a null score from all three
versions of the detector, including the corrected one. The persistence of
the failure after removing the geometric shortcut indicates that its cause
does not lie in the generator.

The properties of the system suggest an explanation. TOI-141~b is a
super-Earth of $1.74\,R_\oplus$ on a 1.008~d orbit around a bright G dwarf
($V = 7.9$), a configuration atypical relative to the dominant population
of the training and validation sets --- Earth-sized planets around M and K
dwarfs. Its period further lies at the upper limit of the
ultra-short-period definition, and the radius ratio against a solar-type
host produces a shallow transit ($\sim$250~ppm) on a star whose brightness
introduces its own instrumental systematics. The combination places this
object in a region of parameter space sparsely represented during training.

We report this case as a known limitation: a confirmed object that the
method systematically rejects, and whose recovery would require broadening
the diversity of the training population towards solar-type hosts.

\subsection{Foreshadowing}\label{sec:fore}

It is worth noting what this sequence reveals, because it prefigures the
central result of this work. Version v2 attained a synthetic AUC of 0.99
while its operational performance collapsed, and the cause was that the
classifier had learned a regularity of the generation procedure rather than
the physics of the problem. \emph{An excellent evaluation metric coexisted
with a useless detector.} \citet{Geirhos2020} note that detecting shortcuts
requires evaluating under conditions different from those of training. The
following section applies that principle systematically, with a
considerably more severe result.

\section{Performance against real false positives}\label{sec:collapse}

\subsection{The benchmarking problem}\label{sec:bench}

A transit classifier trained on synthetic data is often first assessed
against simulated signals or curated eclipsing-binary catalogues, whereas the
leading classifiers are evaluated on real vetted false positives
(Sect.~\ref{sec:intro-dl}). Against a catalogue of this kind, our v3 detector
attains an AUC of 0.97 when discriminating confirmed USPs from the
eclipsing binaries of \citet{Prsa2022}, a catalogue of 4584 systems
identified in sectors 1--26 from two-minute cadence photometry.

We argue that this figure measures a substantially easier problem than the
operational one. The reason is twofold.

First, the nature of the negatives. Catalogued eclipsing binaries
constitute a curated and morphologically distinctive population: they
frequently display secondary eclipses, ellipsoidal variation, V-shaped
profiles, or eclipse depths orders of magnitude larger than a planetary
transit. A convolutional classifier learns to exploit these features
readily. The false positives that actually contaminate a TESS transit
search are of a different and considerably more adverse nature, dominated
by \textit{blends} --- genuine eclipses originating in a source other than
the target and diluted within the photometric aperture --- together with
neighbour contamination, short-period stellar variability, and residual
instrumental systematics.

Second, and critically for TESS, the spatial resolution. The TESS pixel
scale is 21~arcsec \citep{Ricker2015}, against $\sim$4~arcsec for
\textit{Kepler}. This difference, exceeding a factor of five in linear
scale and hence twenty-five in area, implies a far higher density of
contaminating sources per aperture. \citet{Tey2023} attribute the poorer
performance of these classifiers on TESS relative to \textit{Kepler}
precisely to larger pixels, higher false-positive rates, and shorter
observational baselines. A performance figure obtained on \textit{Kepler}
--- where blending is comparatively benign --- does not transfer, and
validation against synthetic EBs does not capture this population at all.

\subsection{A representative test set}\label{sec:testset}

To quantify performance against the real population of contaminants we
constructed an independent test set from the TESS Objects of Interest
catalogue \citep[TOI;][]{Guerrero2021}.

The property that makes this catalogue suitable is not its size but its
provenance. Threshold-crossing events identified by the SPOC
\citep{Jenkins2016} and QLP \citep{Huang2020} pipelines are examined by the
TESS Science Office before receiving a TOI designation, and events
attributable to eclipsing binaries, variable stars, instrumental
systematics or non-planetary transients are excluded from the TOI catalogue
and archived separately. Consequently, TOIs subsequently labelled as false
positives \emph{are not morphologically obvious eclipsing binaries} ---
those were discarded before designation --- but signals that passed both
automated filtering and human vetting as plausible planetary candidates,
and whose non-planetary nature was established only through photometric,
spectroscopic or high-angular-resolution follow-up. They constitute
precisely the population of difficult contaminants that a detector must
learn to reject in operation, and that no eclipsing-binary catalogue
represents.

We selected all TOIs with catalogued orbital period $P < 1.5$~d, obtaining
954 objects. This window is wider than the conventional definition of the
ultra-short-period regime, $P < 1$~d \citep{Winn2018}, and the choice
deserves justification.

First, the one-day cut-off is not astrophysically motivated: as
\citet{PuLai2019} note, it was adopted historically because planets known
below that period were rare at the time, not because it marks a physical
transition. Second, the occurrence rate is discontinuous in its
neighbourhood, with an excess of planets immediately below the cut and
appreciably different slopes of the period distribution on either side
\citep{LeeChiang2017}. Third, and most relevant here, the migration
pathways proposed for the formation of ultra-short-period planets source
them from just above that boundary: from orbital periods of $\sim$1--3~d in
the moderate-eccentricity scenario of \citet{PuLai2019}, and $\sim$1--5~d
in the obliquity-driven tidal runaway of \citet{MillhollandSpalding2020}.
The interval between 1 and 1.5~d is therefore physically continuous with
the population of interest rather than external to it, and truncating the
search exactly at one day would sever a transition region. The planet
validated here has $P = 0.294$~d, comfortably within the
ultra-short-period regime in the strict sense. We adopted as
ground truth the disposition assigned by the TESS Follow-up Observing
Program (TFOP) Working Group:

\begin{itemize}
\item \textbf{Positive class} --- \textit{Confirmed Planet} (CP) and
      \textit{Known Planet} (KP): planets confirmed by independent
      follow-up. After requiring SPOC two-minute cadence photometry,
      $N = 111$.
\item \textbf{Negative class} --- \textit{False Positive} (FP) and
      \textit{False Alarm} (FA): objects rejected by TFOP follow-up. Under
      the same requirement, $N = 253$.
\end{itemize}

We deliberately excluded the \textit{Planet Candidate} (PC) and
\textit{Ambiguous Planet Candidate} (APC) dispositions, which by definition
lack ground truth. These objects take no part in the evaluation and are
reserved as the application set of Sect.~\ref{sec:application}.

We downloaded PDCSAP light curves \citep{Smith2012,Stumpe2012} at
two-minute cadence, processed by the SPOC pipeline, from the Mikulski
Archive for Space Telescopes using \texttt{lightkurve}
\citep{Lightkurve2018}. Each curve was phase-folded at the catalogued
period, rebinned to 256 points, standardised, and cyclically shifted to
place the transit minimum at bin 128. This procedure is identical to that
applied to the synthetic training set, so that any degradation observed is
attributable to the nature of the data and not to a preprocessing mismatch.

Because several distinct samples appear in what follows, we summarise them
in Table~\ref{tab:datasets}. Two are used for evaluation, one for
retraining, one for the blend measurement and one for the search itself;
they should not be confused with one another.

\begin{table*}[t]
\caption{Samples used in this work. The first two are evaluation sets built
from different sources; the third is the application set, disjoint from the
others by construction.}\label{tab:datasets}
\begin{tabular}{@{}llll@{}}
\toprule
Sample & Composition & Size & Used for \\
\midrule
Catalogue benchmark & confirmed USPs (positives)  & 250 &
  conventional evaluation \\
                    & Pr\v{s}a EBs (negatives)    & 263 & (Sect.~\ref{sec:auc}) \\
\addlinespace
TFOP benchmark      & TOI CP/KP (positives)       & 111 &
  realistic evaluation \\
                    & TOI FP/FA (negatives)       & 253 &
  and retraining (Sects.~\ref{sec:auc}, \ref{sec:retrain}) \\
\addlinespace
Blend subsample     & TOI FP/FA with pixel data   & 145 &
  measuring the blend fraction (Sect.~\ref{sec:blends}) \\
\addlinespace
Application set     & TOI PC/APC, no ground truth & 435 &
  the search itself (Sect.~\ref{sec:application}) \\
\botrule
\end{tabular}
\end{table*}

We acknowledge three limitations. First, the positive class carries a
selection bias towards bright objects with deep transits, easier to confirm
through radial velocities. Second, the classes are imbalanced (111 against
253), which is why we adopt the AUC as the primary metric, being invariant
to class proportions. Third, the TFOP disposition, although the community
standard, reflects the state of follow-up at the time of query.

\subsection{The performance collapse}\label{sec:auc}

\begin{figure}[t]
\centering
\includegraphics[width=\textwidth]{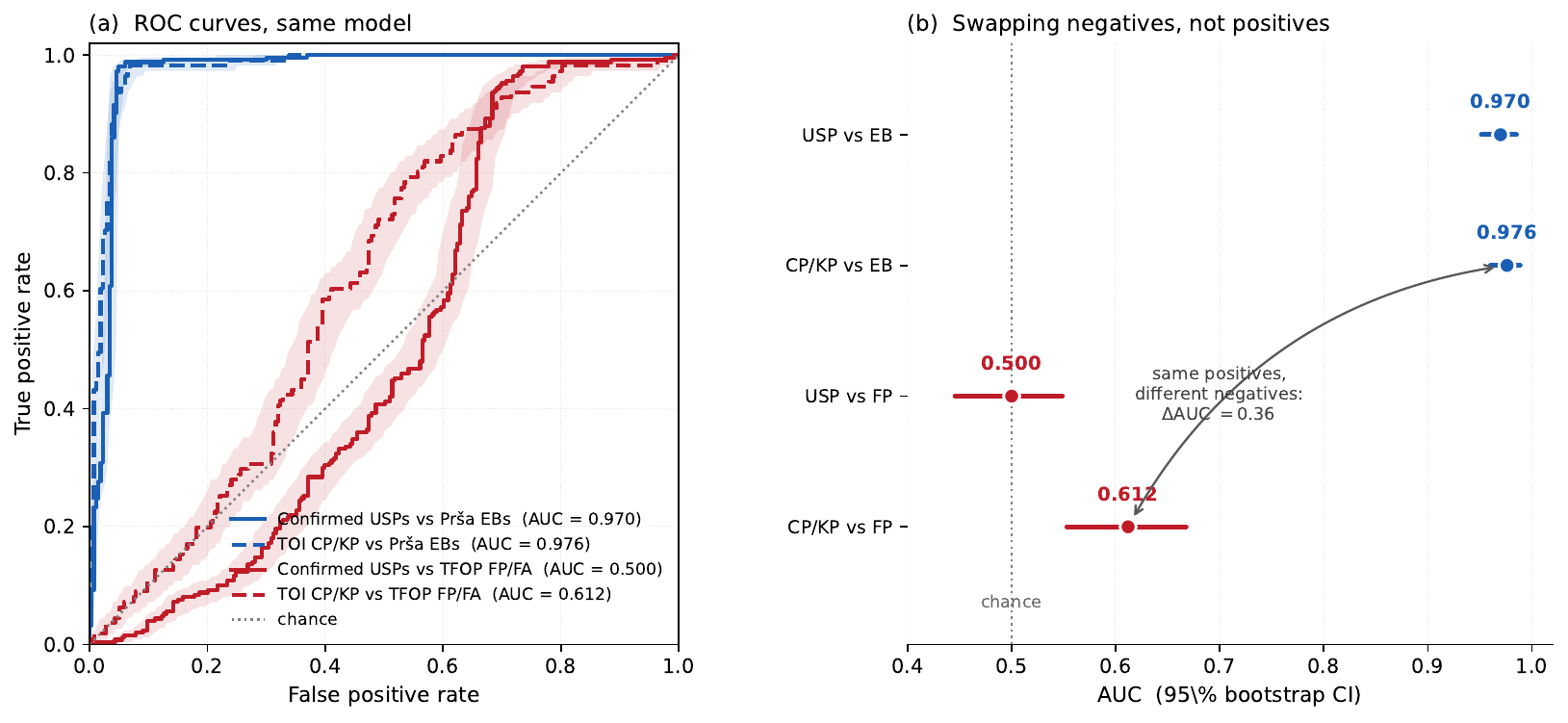}
\caption{The performance collapse. Panel (a): ROC curves for the same v3
model evaluated on all four combinations of two positive and two negative
classes, with 68\% bootstrap confidence bands. Colour encodes the negative
class: blue for catalogued eclipsing binaries \citep{Prsa2022}, red for
real TFOP false positives. Panel (b): corresponding AUC values with 95\%
bootstrap confidence intervals. Holding the positive class fixed and
substituting the negatives drives the AUC from 0.97 to chance, while
substituting the positives leaves it essentially unchanged.}
\label{fig:collapse}
\end{figure}

The v3 detector was initially evaluated on 250 confirmed ultra-short-period
planets against 263 eclipsing binaries from \citet{Prsa2022}, yielding an
AUC of 0.97. Evaluated on the set described above, the value falls to 0.61.
Since the two sets differ simultaneously in their positive and negative
classes, this comparison alone does not attribute the degradation to either
cause. To isolate the responsible variable, we evaluated the same model,
without any retraining, on all four combinations of the two positive and
two negative classes (Fig.~\ref{fig:collapse}). Confidence intervals at the
95\% level were obtained by stratified bootstrap with 2000 resamples.

\begin{table}[t]
\caption{Factorial evaluation of the v3 detector. The same model, without
retraining, is evaluated on all four combinations of positive and negative
classes.}\label{tab:2x2}
\begin{tabular}{@{}llcc@{}}
\toprule
Positives & Negatives & AUC & 95\% CI \\
\midrule
Confirmed USPs (250) & Pr\v{s}a EBs (263) & 0.9699 & [0.951, 0.985] \\
Confirmed USPs (250) & TFOP FP/FA (253) & \textbf{0.5001} & [0.446, 0.549] \\
TOI CP/KP (111) & Pr\v{s}a EBs (263) & 0.9761 & [0.960, 0.989] \\
TOI CP/KP (111) & TFOP FP/FA (253) & 0.6120 & [0.553, 0.668] \\
\botrule
\end{tabular}
\end{table}

The reading by rows and columns is unambiguous. Holding the negative class
fixed, substituting one positive class for the other alters the AUC by
0.006 (0.9699 $\rightarrow$ 0.9761) against the Pr\v{s}a EBs and by 0.112
against the TFOP false positives; in the first case the difference is
negligible and in both the confidence intervals overlap widely. Holding the
positive class fixed, by contrast, substituting catalogued eclipsing
binaries for real TFOP false positives drives the AUC from 0.9699 to
0.5001, with disjoint confidence intervals separated by more than four
tenths.

\emph{The degradation is attributable almost entirely to the nature of the
negative class, and not to that of the positive class.}

The value of 0.5001 deserves emphasis: it is that of a random classifier,
and its confidence interval contains 0.5. Faced with real TFOP false positives,
the v3 detector --- the same that attains 0.97 against catalogued binaries
--- shows no discriminating power. We checked that this result is not driven by differences in period,
which set the duration in phase of the global view. Splitting the TFOP
benchmark at $P = 1$~d, v3 reaches AUC $= 0.634$ below (55 planets, 136
false positives) and 0.591 above (56 planets, 117 false positives), and the
catalogued period alone does not separate the classes (AUC $= 0.47$). The
false positives with $P < 1$~d receive slightly \emph{higher} scores than
those with $1 \le P < 1.5$~d, so restricting the negatives to the strict
ultra-short-period regime would not raise the AUC of 0.50.

All 877 light curves used in the four experiments were downloaded from MAST
with \texttt{lightkurve}, selecting exclusively SPOC-processed PDCSAP
photometry at 120~s cadence, and subjected to the same procedure. The
observed degradation cannot therefore be attributed to differences in
instrumental provenance or preprocessing.

We state explicitly that the two positive classes are not independent: 109
of the 111 TOIs with CP/KP disposition also appear among the 250 confirmed
USPs of the original set. Far from weakening the conclusion, this
reinforces it: with essentially nested positive classes, the variation
observed on changing only the negative class admits no alternative
explanation in terms of sample composition.

\subsection{The irreducible blend fraction}\label{sec:blends}

To identify the origin of the failure we applied a difference-image
analysis \citep{Bryson2013,Twicken2018} to a sample of the TFOP false
positives. We downloaded the target pixel files of each object (SPOC,
120~s cadence) and constructed the difference image by subtracting the
median of the in-transit frames from the median of the out-of-transit
frames, adopting a half-width of 0.05 in phase as the in-transit interval.
The resulting image contains only the flux that disappears during the
eclipse; its flux-weighted centroid locates the source of the signal. We
compared that position with the centroid of the target measured on the
out-of-transit image, and classified as a blend any case with an offset
exceeding 1.0~pixel.

We adopted an absolute threshold rather than a criterion of statistical
significance. In preliminary tests, a 3$\sigma$ criterion erroneously
classified as blends confirmed planets with offsets of barely 0.4~pixels,
since the centroid uncertainty is small when the signal is deep.

We verified the test bidirectionally on a control sample of confirmed
ultra-short-period planets: five of six show offsets below 1.0~pixel and
are correctly classified as on-target, with offsets between 0.38 and
0.98~pixels. The single failure (1.066~pixels) lies marginally above the
threshold.

\begin{figure}[t]
\centering
\includegraphics[width=\textwidth]{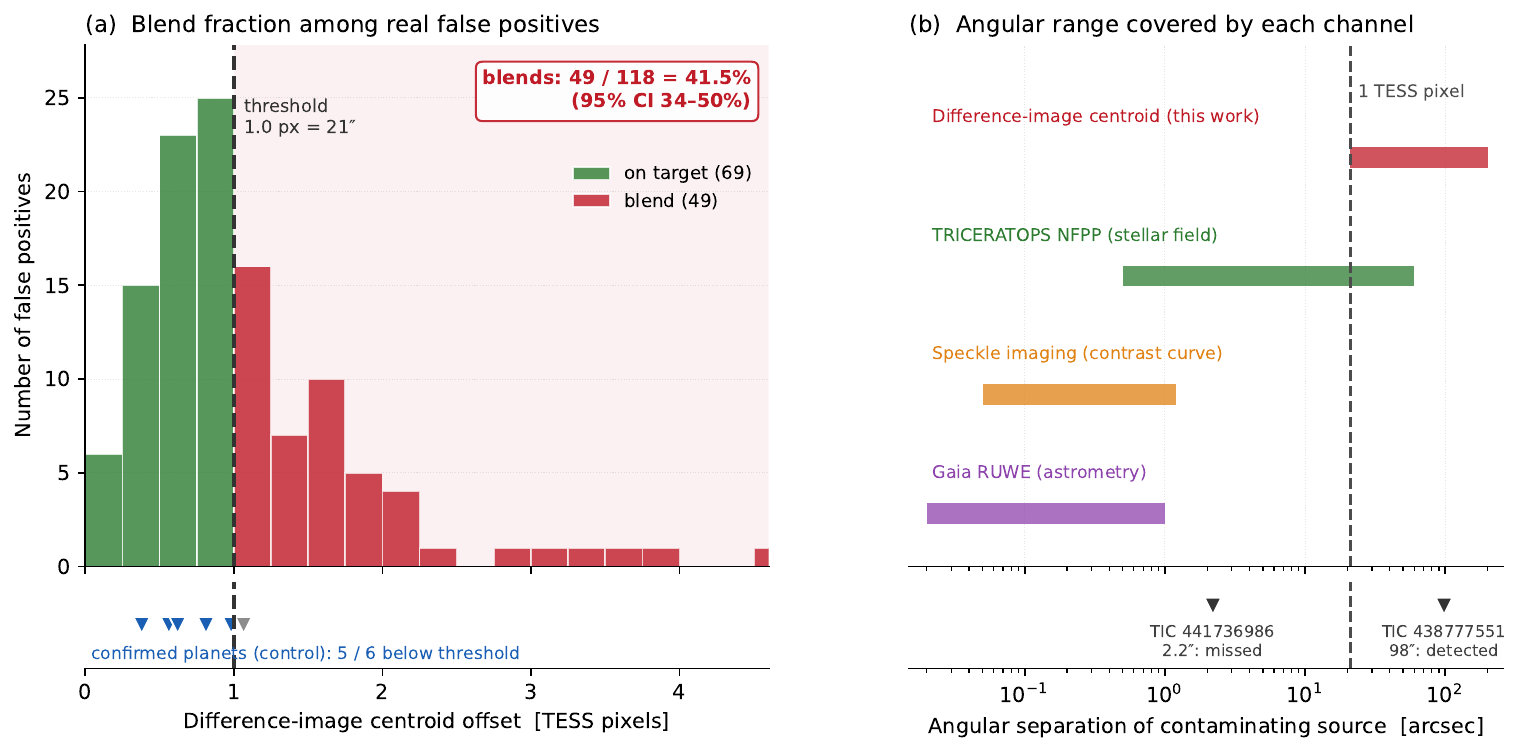}
\caption{The blend fraction and the reach of each vetting channel. Panel
(a): distribution of difference-image centroid offsets for real TFOP false
positives; the dashed line marks the 1.0-pixel decision threshold, and the
markers below show the confirmed ultra-short-period planets used as a
control sample, five of six falling below it. Panel (b): the range of
angular separations over which each channel is sensitive to a contaminating
source. The centroid test reaches only beyond one TESS pixel, which is why
the measured blend fraction is a lower bound: the contaminant of
TIC~441736986, at $2.2\arcsec$, is missed, while that of TIC~438777551, at
$98\arcsec$, is detected.}
\label{fig:centroid}
\end{figure}

We applied the test in two stages. A pilot sample of 30 objects was
analysed first, of which 27 had usable target pixel files and 12 showed a
displaced source. A second, larger pass drew 150 further objects, of which
118 had usable pixel files and 49 showed a displaced source; the remaining
32 lacked two-minute aperture photometry. Combining the two passes,
\textbf{61 of 145 analysed false positives (42.1\%; 95\% Wilson confidence
interval [34.3\%, 50.2\%]) are blends} (Fig.~\ref{fig:centroid}). These
counts refer to false positives drawn from the TFOP benchmark of
Table~\ref{tab:datasets}, and should not be confused with the 250 confirmed
planets that form the positive class of the catalogue benchmark.

For these objects the light curve extracted from the target aperture can
closely resemble that of a shallow planetary transit. The information that
most directly allows correct classification --- the position of
the source --- resides in the pixel data, not in the flux time series. As
\citet{Twicken2018} state in describing this same diagnostic, the technique
exploits spatial information contained in the pixel data, information that
is not available in photometric flux time series. A classifier operating on the folded light curve
alone therefore has, at best, limited ability to resolve this fraction,
although some residual photometric signature may remain
(Sect.~\ref{sec:ordering}).

This fraction is a lower bound, for two reasons. First, the 1.0-pixel
threshold corresponds to 21~arcsec \citep{Ricker2015}; blends produced by
neighbours at smaller separations --- predominant given the TESS pixel
scale --- are not detectable by this procedure. By way of contrast, the
\textit{Kepler} Robovetter operates with thresholds of 1--2~arcsec on
pixel-response-function fits. Second, 32 of the 150 selected objects lacked
two-minute aperture photometry and were excluded; these derive mostly from
faint-star searches in full-frame images, and may be expected to show a
higher, not lower, incidence of blends.

\section{Retraining with real false positives}\label{sec:retrain}

\subsection{Rationale}\label{sec:rationale}

The analysis of Sect.~\ref{sec:blends} establishes that approximately 42\%
of real false positives are blends, difficult to resolve from the light
curve alone. The complementary implication is that the remaining fraction ---
around 58\% --- carries signatures that do reside in the photometry and are
therefore learnable. The collapse documented in Sect.~\ref{sec:auc} is not
due solely to a physical limit: it also reflects that the detector was
never exposed during training to this class of contaminant.

Given the size of the available set (364 curves), training from scratch
would lead to overfitting. We adopt instead a fine-tuning procedure: we
start from the v3 weights --- which already encode transit morphology,
learned from $2 \times 10^5$ synthetic curves --- and readjust the decision
boundary with real data.

This procedure has established precedents in astronomy.
\citet{DominguezSanchez2019} pre-trained a convolutional network on SDSS
galaxy morphologies and fine-tuned it to DES imaging, showing that the
number of labelled examples required is reduced by a factor of a few
relative to training from scratch; equivalent strategies have been employed
between other survey pairs \citep{PerezCarrasco2019,Tang2019}, and
\citet{Walmsley2022} have shown that representations learned by solving a
broad set of morphological tasks substantially improve performance on new
tasks. Our case shares the same structure --- an abundant source of
pre-training examples and a scarce target set --- with the difference that
the source domain is synthetic rather than observational, which aggravates
the distribution shift documented in Sect.~\ref{sec:collapse}.

The direction of transfer conditions the outcome: \citet{Tang2019} observed
that inheriting weights pre-trained on high-resolution imagery improves
performance when fine-tuning on lower-resolution data, whereas transfer in
the opposite direction degrades the classifier. Our case corresponds to the
less favourable direction --- from a synthetic domain, cleaner and of lower
complexity, to an observational one --- which makes the improvement
obtained the more notable.

\subsection{Training set and protocol}\label{sec:protocol}

We use the set described in Sect.~\ref{sec:testset}: 111 TOIs with CP/KP
disposition as positives and 253 with FP/FA disposition as negatives, with
identical photometry and preprocessing. The PC and APC candidates are
excluded by construction, which guarantees that the application set of
Sect.~\ref{sec:application} is disjoint from the training set.

Fine-tuning is performed with the Adam optimiser \citep[][learning rate
$3 \times 10^{-4}$, weight decay $10^{-4}$]{KingmaBa2015}, batches of 32
samples and 40 epochs, minimising binary cross-entropy with positive-class
weighting $w = 253/111 = 2.28$ to compensate the imbalance. The
architecture is inherited unmodified from v3. The learning rate is
deliberately set low: the aim is to readjust the decision boundary without
destroying the representations learned on the synthetic set.

\paragraph{Evaluation protocol.}
We adopt nested stratified five-fold cross-validation. In each fold, 80\%
of the data constitutes the fitting set and the remaining 20\% the test
set. The fitting set is in turn subdivided (80/20, stratified) into
training and inner validation; the stopping epoch is selected by
performance on the inner validation set, and the test fold is evaluated
\emph{once only}, with the weights so selected.

This nesting is not an optional precaution. \citet{VarmaSimon2006}
demonstrated that using the same cross-validation loop to select the model
and to estimate its performance introduces an optimistic bias whose
magnitude grows as the ratio of samples to parameters decreases, and
\citet{CawleyTalbot2010} showed that this bias additionally favours
unnecessarily complex models; both works recommend nested cross-validation
as the remedy. Our set lies precisely in the regime where these authors
predict the strongest effect: 364 samples against of order $5 \times 10^4$
adjustable parameters. We quantify the bias by running both protocols on
the same folds.

\subsection{Results}\label{sec:results-v4}

\begin{table}[t]
\caption{Fold-by-fold cross-validation results. The nested protocol selects
the stopping epoch on an inner validation split and evaluates the test fold
once; the non-nested protocol selects it on the test fold
itself.}\label{tab:cv}
\begin{tabular}{@{}lccc@{}}
\toprule
Fold & v3 & v4 (nested) & v4 (non-nested) \\
\midrule
1 & 0.6649 & 0.8966 & 0.9020 \\
2 & 0.4813 & 0.7201 & 0.7807 \\
3 & 0.6114 & 0.7718 & 0.8226 \\
4 & 0.6791 & 0.8200 & 0.8800 \\
5 & 0.6209 & 0.7945 & 0.8182 \\
\midrule
Mean & $0.6115 \pm 0.0700$ & $\mathbf{0.8006 \pm 0.0582}$ &
       $0.8407 \pm 0.0441$ \\
\botrule
\end{tabular}
\end{table}

Fine-tuning with real negatives raises the AUC from 0.6115 to
$0.8006 \pm 0.0582$, an improvement of $+0.189$ consistent across all five
folds (range $+0.055$ to $+0.239$). The resulting model is designated v4.

The non-nested protocol yields 0.8407, that is, 0.040 above the value
obtained with nesting. The magnitude of the effect agrees with the
qualitative prediction of \citet{VarmaSimon2006} for small sets, and is
comparable to the differences customarily invoked to establish the
superiority of one architecture over another. We report it explicitly
because it illustrates, on our own work, the same class of overestimation
through deficient evaluation that motivates this article.

\subsection{A heuristic bound from the blend fraction}\label{sec:ceiling}

The value attained admits a quantitative interpretation in terms of the
independent measurement of Sect.~\ref{sec:blends}.

The area under the ROC curve admits the interpretation, established by
\citet{HanleyMcNeil1982}, of the probability that a randomly chosen
positive receives a higher score than a randomly chosen negative --- a
quantity equivalent to the non-parametric Wilcoxon statistic. Let $f$ be
the fraction of negatives that are blends. Under the assumption that these
are statistically indistinguishable from positives in the photometric
domain --- so that for such pairs the probability is 0.5 --- and of perfect
separability over the remainder, the maximum attainable AUC is

\begin{equation}
\mathrm{AUC}_{\max} = (1 - f) \cdot 1 + f \cdot 0.5 = 1 - f/2 .
\label{eq:ceiling}
\end{equation}

Substituting the fraction measured in Sect.~\ref{sec:blends},
$f = 0.421$, gives $\mathrm{AUC}_{\max} = 0.789$. Propagating the
confidence interval on $f$ ([0.343, 0.502]) gives a predicted bound in
the range [0.749, 0.829].

\begin{figure}[t]
\centering
\includegraphics[width=\textwidth]{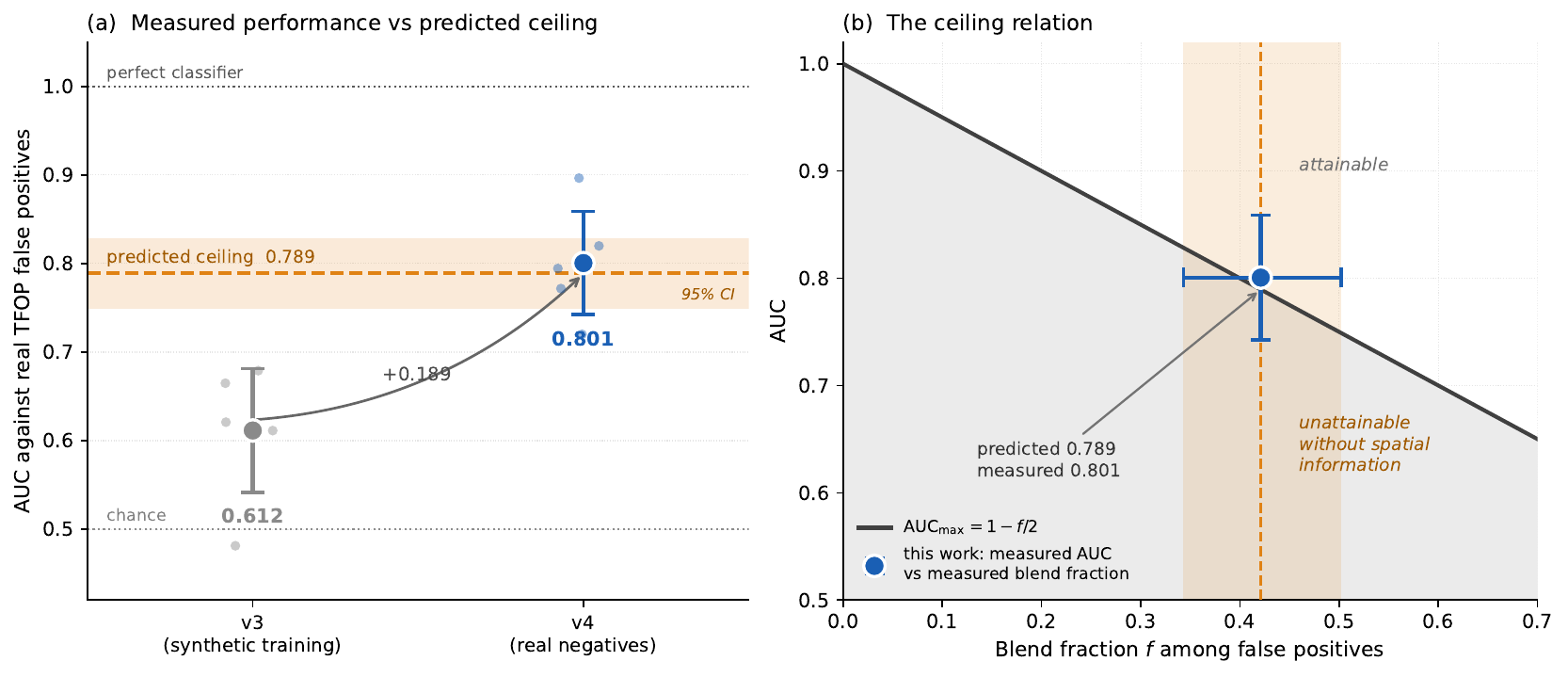}
\caption{Measured AUC compared with the heuristic blend bound. Panel (a): AUC
against real TFOP false positives for the detector trained on synthetic
negatives (v3) and after fine-tuning on real negatives (v4); small points
are individual cross-validation folds, error bars the standard deviation
across folds. The shaded band is the ceiling predicted from the measured
blend fraction, $\mathrm{AUC}_{\max} = 1 - f/2$, with $f = 0.421$
[0.343, 0.502]. Panel (b): the ceiling relation as a function of blend
fraction, with our measurement placed on it with uncertainties on both
axes. The measured value is consistent with the bound; given the uncertainties
and the caveats of Sect.~\ref{sec:ceiling}, this is a consistency check,
not a strong test.}
\label{fig:ceiling}
\end{figure}

The measured value, $0.8006 \pm 0.0582$, lies within that interval and
0.19$\sigma$ from the central prediction (Fig.~\ref{fig:ceiling}).

The measured value is consistent with the bound, but the comparison is not a
strong test: the cross-validation uncertainty ($\pm0.06$) spans most of the
plausible range, and the bound rests on assumptions that our own data
partly contradict (see below).

Two caveats bound this conclusion, and they operate in opposite directions.
The assumption of perfect separability over the non-blends is optimistic,
so that 0.789 constitutes an upper limit. Conversely, not all blends are
wholly indistinguishable: a diluted eclipsing binary may retain a partially
discriminating V-shaped profile, so that the effectively irresolvable
fraction is somewhat smaller than 0.421. Indeed, the classifier score retains some
discriminating power between centroid-identified blends and the surviving
candidates (AUC $= 0.73$; Sect.~\ref{sec:ordering}), showing that blends at
separations above one pixel are not fully indistinguishable in the light
curve. The bound is moreover specific to an input consisting of a single
global view; classifiers with a local transit view or auxiliary inputs are
not subject to it. The agreement should therefore be read as a heuristic
consistency check, not as a determination of a fundamental limit.

\section{Application to unresolved TOIs}\label{sec:application}

\subsection{Application set}\label{sec:appset}

We applied the v4 detector to the set of TOIs with PC or APC disposition
and period $P < 1.5$~d, excluded by construction from the training set for
lacking ground truth. After requiring SPOC 120~s photometry and applying
the preprocessing of Sect.~\ref{sec:testset}, the set comprises 435
objects. These are candidates that have remained unresolved after TFOP
follow-up, and therefore constitute the natural use case for a classifier
intended to prioritise targets.

\subsection{Scoring and centroid filtering}\label{sec:filtering}

The score distribution is strongly bimodal: 140 objects exceed 0.5, 78
exceed 0.8 and 49 exceed 0.9, the threshold we adopt for subsequent
analysis. The threshold is a free parameter of the pipeline, not a
property of the detector: it fixes how many candidates enter the
subsequent stages, and can be lowered or raised according to the vetting
effort available (Sect.~\ref{sec:collier}).

To those 49 we applied the difference-image analysis described in
Sect.~\ref{sec:blends}. Twenty-nine (59\%) show the transit source
displaced by more than one pixel from the target and are classified as
blends, a proportion consistent with the fraction measured on the labelled
false-positive set and with the expectation that the detector is blind to
this class of contaminant. The remaining 20 retain a source coincident with
the target.

We formalise the full procedure in \textbf{CASCADE-USP} (CNN-Assisted
Sequential Centroid And Dilution Evaluation), a pipeline that chains
photometric classification (Sect.~\ref{sec:retrain}) with difference-image
centroid analysis (Sect.~\ref{sec:blends}) and evaluation of dilution by
contaminating sources through TRICERATOPS \citep{Giacalone2021} and
high-angular-resolution contrast curves, which it locates and retrieves
automatically from ExoFOP. The design responds directly to the diagnosis of
this work: since a substantial fraction of false positives cannot be
reliably resolved from the light curve, the pipeline does not pursue a better classifier but
the systematic incorporation of the information channels that photometry
does not contain --- source position through pixel-level analysis, and
stellar field and unresolved companions through scenario modelling with
high-resolution imaging \citep{Scott2021}. Its output is not a verdict but
a decision table which, for each candidate, additionally records
\emph{what information is missing for validation}, allowing observing time
to be prioritised. The framework is independent of the period regime: only
the classification stage is domain-specific, so that adaptation to other
regimes reduces to retraining the classifier on the corresponding
population.

\subsection{Individual vetting}\label{sec:vetting}

Table~\ref{tab:candidates} lists the 20 surviving candidates with their
scores, centroid offsets, Gaia RUWE values and outcome. Each was analysed
individually through TRICERATOPS with contrast curves where available,
complemented by inspection of the TFOP observing notes.

\begin{table*}[t]
\caption{The 20 candidates surviving the centroid filter, ordered by v4
score. Offsets are difference-image centroid displacements (blend threshold
1.0~pixel). RUWE values are from Gaia DR3; entries above 1.4 are flagged as
indicative of an unresolved companion \citep{Lindegren2021}. Dashes denote
sources without a published value.}\label{tab:candidates}
\begin{tabular}{@{}llccccl@{}}
\toprule
TIC & TOI & $P$ (d) & offset (px) & score & RUWE & Outcome \\
\midrule
44797824  & 865.01  & 0.746 & 0.60 & 0.9979 & 1.12 & EB on target (FPP 0.27) \\
465933345 & 6274.01 & 1.111 & 0.67 & 0.9868 & 1.06 & blend (NFPP 1.00) \\
165412561 & ---     & 1.468 & 0.88 & 0.9842 & \textbf{5.41} & chromatic blend \\
160522555 & ---     & 1.307 & 0.78 & 0.9751 & 0.91 & V-shaped, NEBs uncleared \\
469782185 & 1123.01 & 1.209 & 0.60 & 0.9748 & ---  & centroid offset (ExoFOP) \\
409315469 & ---     & 1.075 & 0.70 & 0.9729 & 1.05 & chromatic blend \\
441600335 & 4159    & 0.866 & 0.55 & 0.9723 & ---  & chromatic blend (FPP 0.26) \\
131195882 & 6597    & 0.910 & 0.45 & 0.9720 & 0.98 & no ground-based detection \\
200001699 & ---     & 0.324 & 0.44 & 0.9717 & \textbf{1.49} & blend (NFPP 0.38) \\
119290630 & ---     & 1.488 & 0.90 & 0.9697 & 1.01 & no contrast curve \\
76835246  & 7483    & 0.513 & 0.35 & 0.9688 & 1.06 & no high-resolution imaging \\
301455423 & 3326    & 1.066 & 0.75 & 0.9676 & 1.00 & source on $3.9\arcsec$ neighbour \\
366754985 & ---     & 1.447 & 0.49 & 0.9665 & \textbf{2.52} & neighbours uncleared \\
288671122 & ---     & 0.287 & 0.65 & 0.9618 & 1.05 & blend (NFPP 0.26) \\
\textbf{418343229} & \textbf{6662.01} & \textbf{0.294} & \textbf{0.94}
                   & \textbf{0.9577} & \textbf{1.23} & \textbf{validated planet} \\
200606486 & 6209    & 0.748 & 0.99 & 0.9418 & \textbf{4.25} & chromatic blend \\
90935995  & 6801.01 & 1.224 & 0.63 & 0.9384 & 0.99 & low-contrast EB (FPP 0.07) \\
438777551 & 3161    & 1.401 & 0.89 & 0.9278 & 1.17 & source on $98\arcsec$ neighbour \\
441736986 & 5657.01 & 0.804 & 0.94 & 0.9087 & 0.90 & blend: $2.2\arcsec$ neighbour \\
124453339 & 3106.01 & 1.485 & 0.51 & 0.9005 & \textbf{2.81} & unresolved companion \\
\botrule
\end{tabular}
\end{table*}

\subsection{Score ordering}\label{sec:ordering}

The validated object ranks 26th of 435 by score, and fifteenth among the 20
survivors. The two highest-scoring candidates (0.998 and 0.987) proved to
be, respectively, an eclipsing binary on the target and a blend with
NFPP $= 1$.

Comparing the 20 survivors with the 29 candidates rejected by the centroid
test, the classifier score shows a weak but significant discriminating
ability (AUC $= 0.73$, $p = 0.006$, Mann--Whitney): blends at separations
above one pixel score somewhat lower, suggesting that their dilution leaves
a residual detectable signature. The separation is nevertheless
insufficient for operational use --- the highest-scoring blend exceeds
fifteen of the twenty survivors --- and does not extend to the problem that
matters: within the surviving set, the score \emph{does not order by
planetary likelihood}. This is the expected consequence of the result of
Sect.~\ref{sec:blends}: the upper end of the score distribution
concentrates precisely the cleanest and deepest transit signals, which are
also those producing the most convincing blends and binaries. The useful
discrimination does not come from the detector's ranking but from the
complementary information that the detector cannot see.

\subsection{False-positive taxonomy}\label{sec:taxonomy}

Of the twenty surviving candidates, nineteen fail the validation criteria.
Classifying them by failure mode is more informative than their bare count,
because \emph{each class was resolved through a different information
channel, and none of those channels is present in the folded light curve}.

\paragraph{(a) Chromatic blends --- 4 objects}
(TIC 165412561, 409315469, 441600335, 200606486). Multiband ground-based
photometry reveals a marked dependence of transit depth on wavelength. TIC
165412561 goes from 5.67~ppt in $g$ to 14.92~ppt in $z$; TIC 409315469 from
0.42 to 7.40~ppt between the same bands; TIC 441600335 shows a factor of
forty between $g$ and $z_s$. A planetary transit produces moderate
chromatic variation through limb darkening, never of this order. Such
measurements were obtained with simultaneous multiband instruments such as
MuSCAT2 \citep{Narita2019}; the use of multicolour photometry to validate
or reject ultra-short-period TESS candidates has direct precedent in
\citet{Parviainen2020}. \emph{Channel required: colour.}

\paragraph{(b) Spatial blends --- 7 objects}
(TIC 465933345, 469782185, 200001699, 301455423, 288671122, 438777551,
441736986). The transit source is localised on a star other than the
target, or the target is contaminated by a neighbour that cannot be
separated. TIC 438777551 is the clearest case: ground-based photometry
detects no transit on the target, while a neighbour at 98~arcsec shows a
25~ppt event. TIC 465933345 receives FPP $=$ NFPP $= 1.00$ from
TRICERATOPS.

TIC 441736986 deserves separate mention. Two independent ground-based
detections report a $\sim$7--9~ppt event, but both apertures are
contaminated by a neighbour at 2.2~arcsec with $\Delta T = 2.3$, bright
enough to produce the observed event on its own, and neither observation
can separate the pair: the seeing (5.2~arcsec FWHM) exceeds the separation
by a factor of two. The TFOP notes state explicitly that a focused
observation is still required to determine the source. Crucially, this
object \emph{passed} our centroid test with an offset of 0.94~pixels: at
2.2~arcsec the contaminant lies one tenth of a TESS pixel from the target,
well within the regime identified in Sect.~\ref{sec:blends} as inaccessible
to difference-image centroid analysis. It is a direct illustration of why
the measured blend fraction is a lower bound. \emph{Channel required:
position, stellar field, and angular resolution below the TESS pixel
scale.}

\paragraph{(c) Eclipsing binaries on target --- 2 objects}
(TIC 44797824, 90935995). The signal originates unambiguously on the target
--- TIC 44797824 obtains NFPP $= 0.000$, and follow-up with six MEarth
telescopes \citep{NutzmanCharbonneau2008} confirms the source --- but
statistical modelling favours eclipsing-binary scenarios over the planetary
one. TIC 90935995, with a depth of 27~ppt, would imply a radius of
approximately one Jupiter radius. In both cases the application of
high-angular-resolution contrast curves reduces the FPP without reaching
the validation threshold. \emph{Channel required: scenario modelling plus
high-resolution imaging.}

\paragraph{(d) Irreproducible signal or uncleared field --- 3 objects}
(TIC 131195882, 160522555, 366754985). Ground-based follow-up does not
reproduce the event, or fails to clear the neighbouring stars in the field.
\emph{Channel required: independent photometry.}

\paragraph{(e) Insufficient data --- 2 objects}
(TIC 119290630, 76835246). These lack high-angular-resolution imaging,
without which the FPP remains dominated by unresolved-companion scenarios.
They are not demonstrated false positives but candidates that cannot be
validated with the available information. \emph{Channel required:
high-resolution imaging.}

\paragraph{(f) Astrometric evidence of an unresolved companion --- 1 object}
We queried the Gaia DR3 renormalised unit weight error (RUWE) for all
twenty surviving candidates. The parameter quantifies the goodness of fit
of a single-star astrometric solution \citep{Lindegren2021}: values near
unity indicate a satisfactory solution, while elevated values indicate that
the single-star model describes the data poorly, typically because of an
unresolved companion. Thresholds adopted in the literature range from 1.25
\citep{Penoyre2022} to 1.4 \citep{Lindegren2021}.

Eighteen candidates have a published value, of which \emph{five (28\%)
exceed 1.4}, against the $\sim$12\% characteristic of the \textit{Kepler}
sample: the surviving candidates are enriched in astrometric binarity by
more than a factor of two. No candidate falls in the ambiguous interval
between 1.25 and 1.4, so none of our conclusions depends on the choice of
threshold.

Four of those five had already been rejected through independent channels
--- two by chromaticity, one by uncleared neighbours and one by elevated
NFPP --- so that astrometry independently confirms diagnoses reached by
other routes. The concordance is physically expected in the chromatic
cases: TIC 165412561 (RUWE $=5.41$) and TIC 200606486 (RUWE $=4.25$) are
plausibly perturbed astrometrically by the same companion that produces the
wavelength-dependent dilution --- the same physical object detected through
two independent channels.

The fifth, TIC 124453339, was flagged by astrometry alone. It presented the
cleanest ground-based record of the sample: two independent detections of
consistent depth (4.5 and 4.6~ppt) in uncontaminated apertures, with 21 of
23 neighbours cleared, no chromaticity and no centroid offset. It
nevertheless carries RUWE $=2.81$. Consistently, Gaia publishes no
astrophysical parameters for this source and the TIC lacks both mass and
parallax; TRICERATOPS reports insufficient information to validate it.
Neither the photometry, nor the centroid analysis, nor the contrast curve
flagged this object; the astrometric solution did.

The validated planet lies in the clean group (RUWE $=1.23$), providing an
additional independent line of support for its planetary nature.

We note that the channel is complementary rather than redundant: RUWE is
sensitive to companions below the Gaia resolution limit, whereas resolved
neighbours at several arcseconds --- such as the contaminant $2.2\arcsec$
from TIC 441736986, whose RUWE is 0.90 --- require stellar-field modelling
instead.

\medskip
The central observation of this section is the following: \emph{none of the
nineteen false positives was distinguishable from a planetary transit on
the basis of the folded TESS light curve.} All had passed automated
filtering, the human vetting of the TESS Science Office, and our classifier
with a score above 0.9; sixteen additionally passed the centroid test.
Their nature was established through information on colour, position, the
surrounding stellar field, independent photometry, or astrometry ---
channels all absent from the flux time series.

This carries a practical consequence beyond the specific case: \emph{no
single complementary test is sufficient}. TRICERATOPS resolves classes (b)
and (c) but is blind to colour; multiband photometry resolves (a) but does
not quantify binary scenarios; centroid analysis resolves blends at
separations above one pixel but not closer ones; astrometry resolves
unresolved companions but not resolved neighbours. Validating an
ultra-short-period candidate in TESS requires the concurrence of several
independent channels, and the contribution of a photometric classifier is
necessarily limited to prioritisation.

\section{TOI-6662~b: an independent validation}\label{sec:planet}

\subsection{Statistical validation}\label{sec:validation}

TOI-6662~b (TOI-6662.01, TIC 418343229) received a score of 0.958 from the v4 detector
and passed the difference-image test with an offset of 0.94~pixels,
consistent with an on-target origin. Its catalogued period is
$P = 0.2939493$~d (7.05~h), and the host is an M dwarf of radius
$R_\star = 0.332 \pm 0.010\,R_\odot$ and effective temperature
$T_{\rm eff} = 3359 \pm 157$~K, observed by TESS in sectors 19, 20, 26 and
40.

\begin{figure*}[t]
\centering
\includegraphics[width=\textwidth]{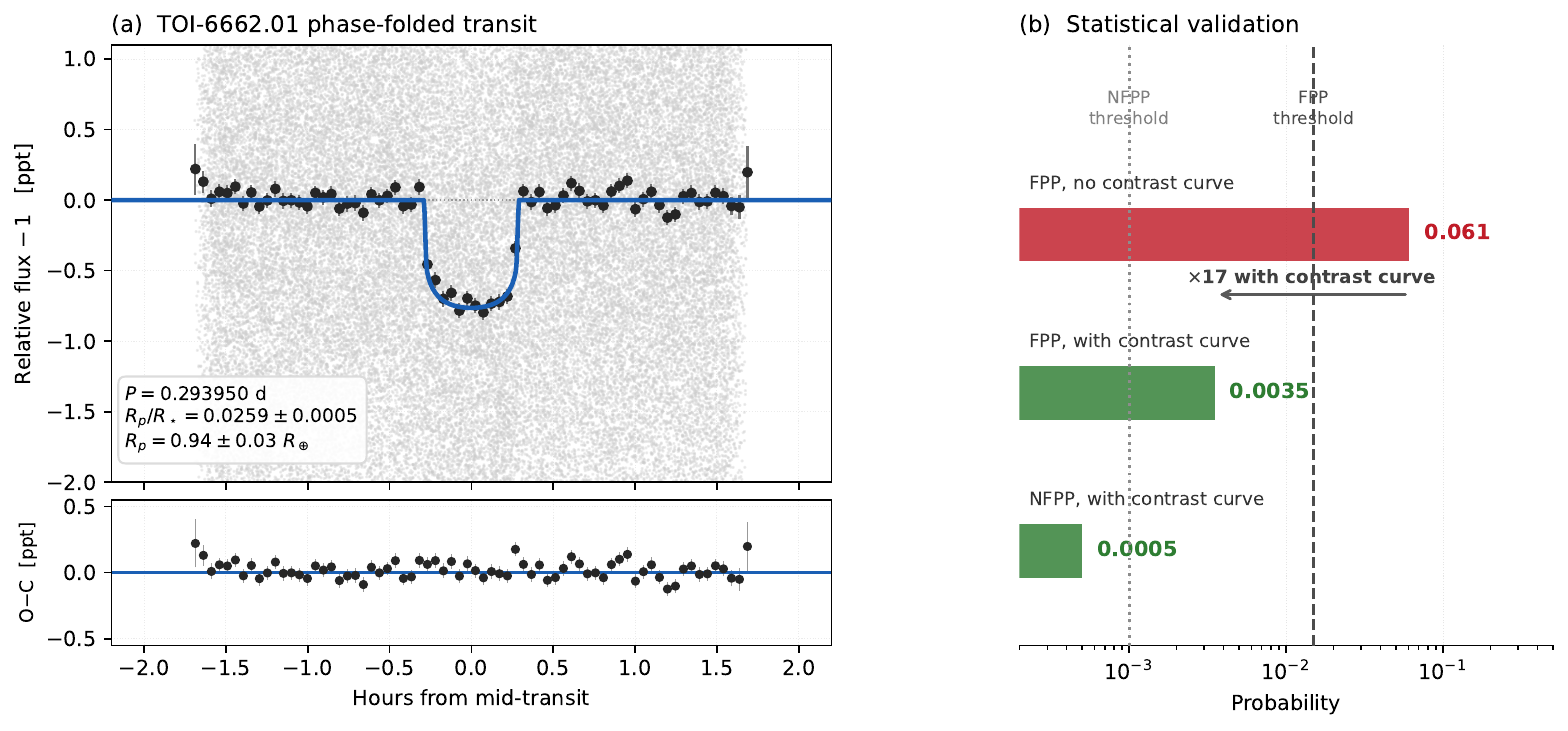}
\caption{TOI-6662~b. Panel (a): phase-folded photometry from sectors 19,
20, 26 and 40 (grey), binned means with standard errors (black), and the
best-fit \texttt{juliet} transit model (blue); residuals below. Panel (b):
false-positive probabilities from TRICERATOPS with and without the Gemini
832~nm speckle contrast curve, against the validation thresholds of
\citet{Giacalone2021}. Incorporating the contrast curve lowers the FPP by a
factor of 17 and moves the candidate across the validation
threshold.}\label{fig:toi6662}
\end{figure*}

We applied TRICERATOPS \citep{Giacalone2021} to the folded and binned light
curve, with the stellar field drawn from TRILEGAL and fifteen independent
realisations of $5 \times 10^5$ samples each. Without high-resolution
imaging information, the false-positive probability is FPP $=0.061$,
placing the object in the \textit{likely planet} category but short of the
validation threshold. Incorporating the contrast curve obtained with the
Gemini speckle interferometer in the 832~nm band \citep{Scott2021}, the
values drop to FPP $= 0.0035$ and NFPP $= 0.00050$ across all fifteen
realisations (Fig.~\ref{fig:toi6662}). Both satisfy the validation criteria
of \citet{Giacalone2021}, which require FPP $< 0.015$ and
NFPP $< 10^{-3}$. \textbf{TOI-6662~b is therefore statistically validated
as a planet.}

The factor of seventeen reduction in FPP on incorporating the contrast
curve illustrates the point developed in Sect.~\ref{sec:taxonomy}:
high-angular-resolution imaging information, absent from the photometry, is
what resolves the unresolved-companion scenarios that dominated the
residual probability.

A further channel supports the validation: TOI-6662~b has RUWE $= 1.23$ in
Gaia DR3, below the thresholds conventionally adopted as indicative of an
anomalous astrometric solution \citep{Lindegren2021,Penoyre2022}, with no
evidence therefore of an unresolved companion.

The target benefits from unusually extensive high-resolution imaging
coverage: ExoFOP records six speckle observations from four instruments ---
Gemini/'Alopeke at 562 and 832~nm, WIYN/NESSI at the same two bands, and
two epochs with the SAI-2.5m Speckle Polarimeter in $I$ --- with contrast
limits between $\Delta 4.5$ and $\Delta 6.8$~mag. The validation adopts the
deepest of these, but the concordance of four independent instruments
strengthens the exclusion of resolved companions.

\subsection{Consistency with TFOP follow-up}\label{sec:tfop}

The statistical validation agrees with independent follow-up. The object
has five ground-based photometric detections of the event, all in
uncontaminated apertures and consistent in time and duration with the
ephemeris. The TFOP notes state that no further observations are required
beyond extreme-precision photometry aimed at characterising transit-timing
variations, which reflects that the follow-up programme already treats the
object as planetary. The apparent chromatic dependence of the depth is
accounted for by the faintness of the M-dwarf host in the blue.

\subsection{Comparison with an independent validation}\label{sec:collier}

After the submission of this work, \citet{Collier2026} published a
statistical validation of the same object, designated TOI-6662~b, within
an automated TRICERATOPS analysis of 443 TESS candidates designed to
measure the contribution of high-resolution imaging to validation. The two
analyses were carried out independently and reached the object by
different routes. In \citet{Collier2026} it entered through a target list
drawn from TFOP planet candidates in predefined categories of parameter
space; here it was recovered from a pool of 435 unresolved TOIs by
classifier prioritisation followed by spatial vetting, which is precisely
the task CASCADE was designed for. Table~\ref{tab:collier} compares the
quantities the two works have in common.

\begin{table}[h]
\caption{TOI-6662~b in this work and in \citet{Collier2026}. FPP values
are means over independent TRICERATOPS realisations (15 of
$5\times10^5$ samples here; 20 of $10^6$ there). FPP$_{\rm noCC}$ is the
value obtained without the contrast curve. Masses are estimated from the
mass--radius relation of \citet{ChenKipping2017} in both works.}
\label{tab:collier}
\begin{tabular}{@{}lcc@{}}
\toprule
Quantity & This work & \citet{Collier2026} \\
\midrule
$P$ (d) & 0.293950 & $0.29395013^{+0.00000005}_{-0.00000006}$ \\
$R_p/R_\star$ & $0.0259 \pm 0.0005$ & $0.0260^{+0.0005}_{-0.0006}$ \\
$a/R_\star$ & $3.81 \pm 0.14$ & $3.85^{+0.12}_{-0.14}$ \\
$R_p$ ($R_\oplus$) & $0.94 \pm 0.03$ & $0.94^{+0.03}_{-0.04}$ \\
$M_p$ ($M_\oplus$, estimated) & $\approx 0.8$ & 0.78 \\
FPP (with contrast curve) & $3.5\times10^{-3}$ & $1.6\times10^{-11}$ \\
FPP$_{\rm noCC}$ & 0.061 & 0.026 \\
NFPP & $5.0\times10^{-4}$ & $<10^{-3}$ \\
Contrast curve & 'Alopeke, 832~nm & 'Alopeke, 832~nm \\
Gaia DR3 RUWE & 1.23 & $\leq 1.4$ (cut) \\
Centroid test & 0.94~px offset & --- \\
\botrule
\end{tabular}
\end{table}

The transit geometry agrees within one standard deviation in radius
ratio, scaled semi-major axis and planetary radius. The classification
agrees as well, and so does the role of the imaging: both analyses find
the object short of the validation threshold without the contrast curve
($\mathrm{FPP}_{\rm noCC} = 0.061$ here, $0.026$ there) and well inside it
with the curve. The absolute FPP values with the curve differ by several
orders of magnitude, which is expected at this level: once the competing
scenarios are suppressed, the residual FPP depends sensitively on light
curve preparation, ephemeris and sampling, and what is robust is the
classification, not the figure. Our fitted period, $P = 0.293950$~d
(Fig.~\ref{fig:toi6662}), coincides with theirs to the precision quoted;
the catalogued value used to set up the analysis, $P = 0.2939493$~d,
differs from it by $8\times10^{-7}$~d, which accumulates to about two
minutes over the baseline from sector 19 to sector 40, small against a
transit duration of $\sim$35 minutes.

The agreement must not be overstated. Both analyses rest on SPOC
photometry, on TRICERATOPS with its priors, and on a Gemini/'Alopeke
contrast curve at 832~nm; they share their principal inputs. What the
comparison establishes is that the validation is reproducible under two
independent implementations, not that it rests on two independent bodies
of evidence. The elements of our analysis without a counterpart in
theirs --- the difference-image centroid test, the review of six speckle
observations from four instruments, and the five ground-based detections
in uncontaminated apertures --- add to the case, while both apply the
Gaia astrometric cut.

For CASCADE the comparison has a specific value. A candidate prioritised
by the classifier, filtered by its centroid and resolved by scene
modelling reaches the same verdict as a dedicated validation effort with
access to the same follow-up. It is the closest available approximation
to an end-to-end external test of the pipeline on a positive case, and it
is the natural end point of the procedure described in this work.

The comparison also permits a test in the opposite direction.
\citet{Collier2026} validate two further ultra-short-period planets within
our period range, TOI-1435~b ($P = 0.686$~d) and TOI-1806~c
($P = 0.502$~d). Both belong to our application set of 435 TOIs, and
neither reached the centroid stage: the v4 detector assigned them scores
of 0.29 and 0.10, ranking them 200th and 274th. Of the three planets in
the application set validated by \citet{Collier2026}, CASCADE therefore
recovers one at its operating threshold of 0.9. These are false negatives
of the classifier, not of the spatial filters, so they leave the blend
fraction of Sect.~\ref{sec:blends} unaffected; but they measure, on real
planets, the cost of a threshold chosen for purity, and they agree with
the conclusion of Sect.~\ref{sec:ordering} that the score orders
candidates for follow-up rather than by planetary likelihood. Both
signals are less favourable for single-band detection than that of
TOI-6662~b: TOI-1435~b has a transit depth of $\sim$150~ppm against
$\sim$680~ppm, and TOI-1806~c orbits a host nearly two magnitudes fainter
($T = 12.6$ against 10.9; \citealt{Collier2026}).

Both would have been recoverable at a lower threshold, at a price that can
be stated exactly. Lowering it to 0.29 would have admitted TOI-1435~b,
with 201 candidates entering the centroid stage instead of 49; admitting
TOI-1806~c as well would have required 0.10, and 274 candidates, 63\% of
the set. Since the classifier costs nothing to run, while the stages after
it --- the centroid analysis and, above all, the individual vetting of the
survivors --- are where the effort lies, the threshold should be set by
the follow-up budget available rather than fixed in advance: a survey
seeking purity with limited resources operates high, one seeking
completeness operates low and accepts a proportionally larger vetting
load. With only three validated planets in the application set, we do
not attempt to calibrate a completeness curve as a function of the
threshold; the figures above illustrate the trade-off rather than measure
it.

\subsection{Transit fit}\label{sec:fit}

We fitted the transit with \texttt{juliet} \citep{Espinoza2019juliet},
which models the transit through \texttt{batman} \citep{Kreidberg2015} and
explores the parameter space by nested sampling with \texttt{dynesty}
\citep{Speagle2020}. To keep the problem tractable given the $\sim$350
transits present in four sectors, we extracted windows of $\pm 3$ transit
durations centred on each predicted event and normalised each window by its
local out-of-transit median.

We fitted period, epoch, radius ratio, impact parameter and stellar density
--- the last with a Gaussian prior derived from the TIC parameters, which
breaks the degeneracy between impact parameter and radius ratio --- with
quadratic limb darkening in the parametrisation of \citet{Kipping2013} and
a white-noise jitter term. The eccentricity was fixed to zero, a choice
justified in Sect.~\ref{sec:tides}.

We obtain $R_p/R_\star = 0.0259 \pm 0.0005$ and
$b = 0.28^{+0.12}_{-0.17}$. Combined with
$R_\star = 0.332 \pm 0.010\,R_\odot$, this yields
$R_p = 0.94 \pm 0.03\,R_\oplus$, with the uncertainty dominated by that of
the stellar radius. TOI-6662~b is therefore an Earth-sized planet.

\subsection{Tidal environment}\label{sec:tides}

With a semi-major axis of $a/R_\star = 3.81 \pm 0.14$, TOI-6662~b orbits
at $1.19 \pm 0.01$ times the tidal disruption distance (fluid Roche limit).
This proximity imposes a minimum bulk density on the planet --- close to
the terrestrial value --- below which it would have been disrupted at this
separation, the minimum-density argument established by
\citet{Rappaport2013} for the 4.2~h planet KOI-1843.03.

Adopting $M_p \approx 0.8\,M_\oplus$ estimated from the mass--radius
relation of \citet{ChenKipping2017}, the orbital decay timescale due to
stellar tidal dissipation, in the constant-lag model
\citep{GoldreichSoter1966}, is approximately $20\,(Q'_\star/10^7)$~Gyr. For
the limit $Q'_\star \gtrsim 10^7$ inferred at the population level for
ultra-short-period planets by \citet{HamerSchlaufman2020}, the orbit is
stable against tidal inspiral on timescales comparable to the age of the
system. At this period tidal circularisation is essentially instantaneous
in astronomical terms, which supports the assumption of a circular orbit
adopted in the fit.

\subsection{Physical characterisation}\label{sec:physical}

The planet receives $366^{+79}_{-67}$ times the terrestrial irradiation.
Its equilibrium temperature, assuming zero albedo and full heat
redistribution ($f = 1/4$), is $\sim$1220~K, rising to $\sim$1555~K on the
dayside in the absence of redistribution
\citep[$f = 2/3$;][]{LopezMoralesSeager2007}. These values place
TOI-6662~b at the boundary of the regime where a surface magma ocean can
be sustained: comparable dayside temperatures have been considered
insufficient for that purpose \citep{ElkinsTanton2012,Srivastava2026}. It
is therefore not unambiguously a lava world, but a body for which the state
of the mantle depends on factors beyond the irradiation alone --- in
particular on any residual tidal heating --- which makes it an object of
interest for future phase-curve measurements. \citet{Collier2026} report
an emission spectroscopy metric of 12.0, the highest among the
terrestrial planets they newly validate, above the threshold of 7.5
recommended by \citet{Kempton2018} for JWST follow-up.

With $R_p = 0.94\,R_\oplus$ and an irradiation far above 100~$S_\oplus$,
the planet lies unambiguously in the lower rocky peak of the radius valley
\citep{Fulton2017}, consistent with a core stripped of its envelope by
photoevaporation --- the ``hot super-Earth desert'' described by
\citet{Lundkvist2016}.

\subsection{Context}\label{sec:context}

\begin{figure*}[t]
\centering
\includegraphics[width=0.92\textwidth]{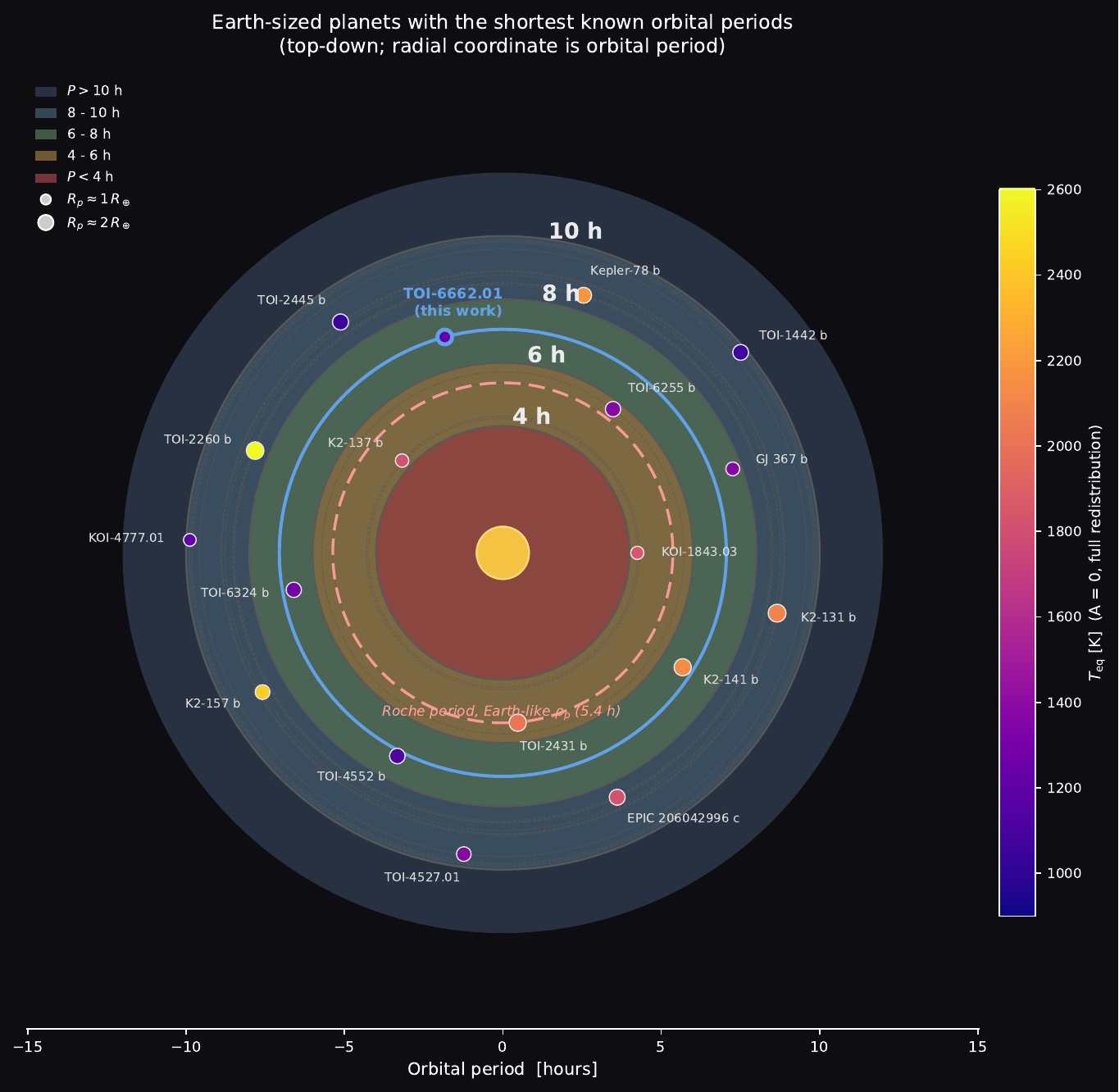}
\caption{Earth-sized planets with the shortest known orbital periods,
viewed top-down with the radial coordinate set by orbital period. Coloured
bands mark period intervals; the dashed ring indicates the Roche-limiting
period for an Earth-like bulk density (5.4~h; \citealt{Rappaport2013}).
Symbol colour encodes equilibrium temperature (zero albedo, full
redistribution) and symbol size scales with planetary radius. TOI-6662~b,
validated independently in this work and by \citet{Collier2026}, is
highlighted.}\label{fig:periods}
\end{figure*}

\begin{figure*}[t]
\centering
\includegraphics[width=0.92\textwidth]{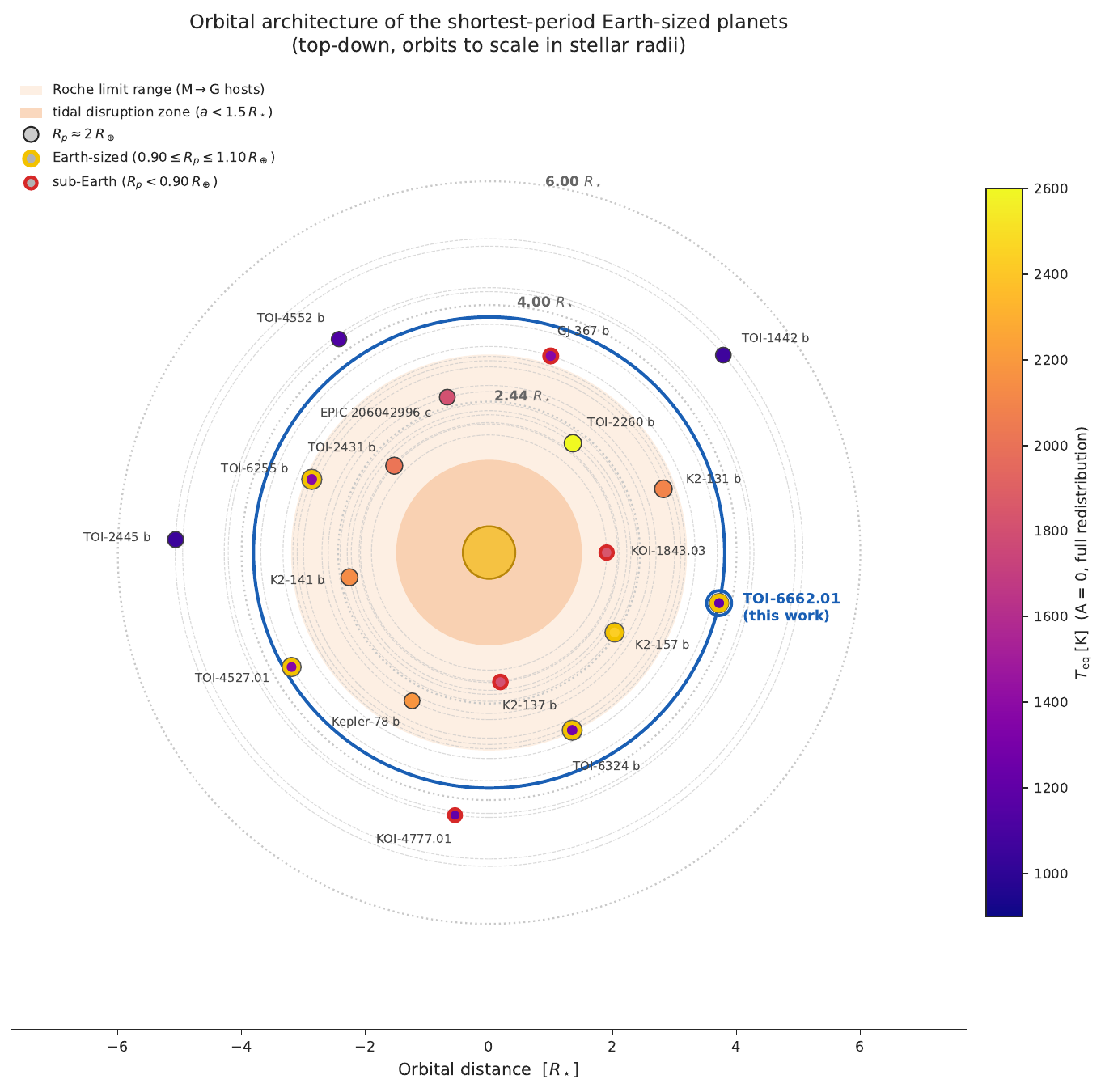}
\caption{Orbital architecture of the same sample, with orbits to scale in
units of stellar radius. Shaded regions indicate the range spanned by the
Roche limit across host spectral types and the inner tidal disruption zone.
Earth-sized planets ($0.90 \leq R_p \leq 1.10\,R_\oplus$) are outlined in
yellow and sub-Earths ($R_p < 0.90\,R_\oplus$) in
red.}\label{fig:architecture}
\end{figure*}

At $P = 0.294$~d, TOI-6662~b falls comfortably within the
ultra-short-period regime, defined by $P < 1$~d \citep{Winn2018}. It is not
a record: the shortest-period confirmed planet is K2-137~b, at 0.18~d
around a $0.463\,M_\odot$ M dwarf \citep{Smith2018}, and the candidate
KOI-1843.03 has a still shorter period \citep{Rappaport2013}. Among
Earth-sized planets with $P<0.42$~d it ranks seventh of eighteen by period
and is the smallest of the four M-dwarf hosts with comparable periods
(Figs.~\ref{fig:periods} and \ref{fig:architecture}).

Its closest analogues are TOI-6255~b (0.238~d, $1.08\,R_\oplus$;
\citealt{Dai2024}), TOI-6324~b (0.279~d, $1.06\,R_\oplus$;
\citealt{Lee2025}) and TOI-4552~b (0.301~d, $1.11 \pm 0.04\,R_\oplus$,
$\rho_p = 7.74 \pm 2.12$~g~cm$^{-3}$, M4.5V host at 27.26~pc;
\citealt{Srivastava2026}), all Earth-sized planets around M dwarfs with
periods within 20\% of ours; its radius places it alongside the dense
sub-Earths such as GJ~367~b \citep{Lam2021}, and its proximity to the Roche
limit relates it to TOI-6255~b, described as a planet on the verge of tidal
disruption. In all these cases, however, the planetary nature was
established through radial velocities or high-angular-resolution imaging
with a mass determination. Only around a dozen ultra-short-period rocky
planets orbiting M dwarfs have both mass and radius determined
\citep{Srivastava2026}; TOI-6662~b is not among them.

Its interest lies not, therefore, in a superior characterisation relative
to its peers, but in the route by which it was identified: the recovery of
an unresolved candidate from the TOI catalogue through automated
prioritisation. In a regime where the occurrence rate of terrestrial USPs
around M dwarfs peaks \citep{SanchisOjeda2014} but the distinction from
false positives is hardest, this is precisely the kind of object that an
carefully characterised classifier can help to rescue --- provided its
output is read as what it is: a prioritisation, not a decision.

We emphasise that this is a statistical validation, not a dynamical
confirmation. The mass remains unmeasured, and definitive confirmation
would require radial velocities or the analysis of transit-timing
variations currently pursued by TFOP. The expected radial-velocity
semi-amplitude, $\approx 1.6$~m~s$^{-1}$ for the inferred mass, is
accessible to high-stability near-infrared spectrographs given the
brightness of the host ($T = 10.9$).

\section{Discussion}\label{sec:discussion}

\subsection{The limits of light-curve-only classification}\label{sec:limit}

A recurring theme of this work is that light-curve information alone is
insufficient for a substantial fraction of real false positives. On the one hand,
difference-image analysis establishes that 42\% of real false positives are
blends (Sect.~\ref{sec:blends}); under the assumption that these are
indistinguishable from transits in the light curve, the maximum attainable
AUC is $1 - f/2 = 0.789$. On the other, retraining with real negatives,
evaluated under nested cross-validation, reaches $0.80 \pm 0.06$
(Sect.~\ref{sec:ceiling}). The two values are consistent, but given the uncertainties and the
caveats of Sect.~\ref{sec:ceiling}, this is a consistency check rather than
proof that our detector has extracted all the information available in the
light curve. A better architecture (with a local view) and a physically
realistic training set could plausibly do better.

The broader point, well established in the field, is that further gains lie in incorporating the information
channels that do distinguish a blend from a planet: colour, through
multiband photometry; position, through centroid analysis or
high-resolution imaging; the stellar field, through scene modelling; and,
as this work adds, astrometry, through the quality of the single-star
solution.

This conclusion is not speculative: it reflects the direction the field has
itself taken. ExoMiner \citep{Valizadegan2022} incorporates centroid motion
data among its inputs; \citet{TardugnoPoleo2024} add background flux and
centroid information to a convolutional classifier, identifying a median
18\% of contaminants; and the TRICERATOPS+ extension
\citep{GomezBarrientos2025} adds ground-based multiband photometry to the
validation framework, integrating precisely the colour channel whose
absence from the light curve we identify as responsible for the irreducible
fraction of chromatic blends. Our results add a measurement of how
large the blend fraction is among real short-period TESS false positives,
and a concrete illustration of each channel's contribution.

\subsection{Implications for detector benchmarking}\label{sec:implications}

Our results illustrate the risk of evaluating a transit detector only
against synthetic or catalogued eclipsing binaries. The same detector that attains an AUC of 0.97 against the
binaries of \citet{Prsa2022} obtains 0.50 --- pure chance --- against the
real false positives of TFOP with an identical positive class
(Sect.~\ref{sec:auc}). The difference is not attributable to the
composition of the positive sample, which we held fixed. This does not apply
to the leading classifiers cited in Sect.~\ref{sec:intro-dl}, which were
trained and evaluated on real vetted false positives; it applies to
detectors, such as our v3, whose evaluation relies on catalogued or
simplistic synthetic binaries. For such detectors, we recommend reporting
performance also against real TFOP false positives, which represent the
population that will determine the false-positive rate in a real search.

The subtlety of the problem is illustrated within our own work. In
retraining the detector, a cross-validation protocol selecting the stopping
epoch by performance on the test fold --- an apparently innocuous practice
--- inflated the AUC by 0.040 (Sect.~\ref{sec:results-v4}), a magnitude
comparable to those customarily invoked to establish the superiority of one
architecture over another. Evaluation bias is not a defect of others: it is
a standing temptation that only a strict protocol --- nested
cross-validation, representative test sets --- keeps at bay.

We add a practical recommendation. The Gaia RUWE query is the cheapest of
the channels employed here: a single call to a catalogue service, of order
one second per target, against minutes for scenario modelling and days to
months for ground-based follow-up. Applied on its own to our sample it
would have flagged five of the nineteen false positives (26\%) with no
observational effort whatsoever. We recommend its inclusion as an early
filter in any candidate-vetting pipeline.

\subsection{Limitations}\label{sec:limitations}

We enumerate the limitations of this work, some of which bound the scope of
the preceding conclusions.

The blend fraction of 42\% is a lower bound, not a complete measurement:
our centroid test, with a threshold of one pixel (21~arcsec), is blind to
blends produced by neighbours at smaller separations, which TRICERATOPS
does capture through the NFPP but which escape population-level
characterisation. TIC 441736986 illustrates the point directly: its
contaminant lies 2.2~arcsec away, one tenth of a TESS pixel, and the object
passed our test with an offset of 0.94~pixels. If closer blends were
included, $f$ would be larger and the heuristic bound correspondingly lower
than 0.789, under the same simplifying assumptions.

The detector itself is deliberately simple and is not representative of
the state of the art. Its input is a single global view of 256 phase bins,
without a local transit view or auxiliary inputs (centroids, stellar or
\textit{Gaia} parameters), and its synthetic training set used trapezoidal
profiles on Gaussian noise, without realistic stellar or instrumental noise,
physical eclipsing-binary models, blended scenarios or physically motivated
parameter distributions, unlike the realistic simulations used by current
pipelines \citep{Osborn2020,Lafarga2026}. The v3 generator, which forced every synthetic
binary to show a secondary eclipse or ellipsoidal modulation, represents
only a narrow subset of the binary population. Conclusions drawn from its
performance apply to this detector, not to transit classifiers in general.

The retraining set is small (364 curves), which lends the AUC figure an
appreciable uncertainty ($\pm 0.06$) and makes the procedure sensitive to
the evaluation protocol. The positive class further inherits the selection
bias of the confirmed catalogue towards bright objects with deep transits.

The detector does not reach the shortest-period regime. The transit search
algorithm used in the survey phase presents an effective floor around
$P \approx 0.6$~d on two-minute photometry, so that planets of shorter
period --- including the most extreme, such as analogues of KOI-1843.03 ---
lie beyond its reach. The validation of TOI-6662~b ($P = 0.294$~d) was
possible because the object already appeared in the TOI catalogue, not
because our survey could have recovered it.

Finally, the validation of TOI-6662~b is statistical: its mass has not
been measured, and its planetary nature, although established with high
confidence through the concurrence of FPP, centroid analysis, contrast
curve, astrometry and ground-based follow-up, does not amount to a
dynamical confirmation. Nor is the pipeline complete at its operating
threshold: two planets of the application set validated by
\citet{Collier2026} scored below 0.9 and were not recovered
(Sect.~\ref{sec:collier}). We did not conduct a systematic search for
additional planets in the system; the detection of companions on wider
orbits is left for future work.

\subsection{TOI-6662~b in context}\label{sec:planet-context}

The validation procedure employed here --- photometric classification,
centroid analysis, TRICERATOPS with a speckle contrast curve, and
confirmation through multiband ground-based photometry --- reproduces that
recently applied to TOI-4552~b \citep{Srivastava2026}. In that case,
however, as in those of its other analogues, the characterisation was
completed with a mass determination from radial velocities.

Its interest lies not, therefore, in a superior characterisation, but in
the route by which it was identified: the recovery of an unresolved
candidate from the TOI catalogue through automated prioritisation. In a
regime where the occurrence rate of terrestrial USPs around M dwarfs peaks
\citep{SanchisOjeda2014} but the distinction from false positives is
hardest, this is precisely the kind of object that an carefully
characterised classifier can help to rescue. That the object was reached
independently by a dedicated validation programme \citep{Collier2026}
does not diminish this point; it confirms that the route through CASCADE
leads to the right answer.

The framework of CASCADE is independent of the period regime: the centroid
analysis and dilution evaluation stages operate on any transit candidate.
Only the classification stage is domain-specific, so that adaptation to
other regimes reduces to retraining the classifier on the corresponding
population.

\section{Conclusions}\label{sec:conclusions}

We have presented the construction, characterisation and application
of a simple convolutional transit detector for ultra-short-period planets in TESS.
Our conclusions are as follows.

\begin{enumerate}

\item For our detector, performance against catalogued eclipsing binaries
did not predict performance against real false positives. Our
detector attains an AUC of 0.97 against the binaries of \citet{Prsa2022}
and \textbf{0.50 --- indistinguishable from chance --- against the real
false positives of TFOP}, with an identical positive class. A factorial
design varying each class separately attributes the degradation almost
entirely to the nature of the negatives (Sect.~\ref{sec:auc}).

\item Blends are a major contributor.
Difference-image analysis of 145 real false positives shows that
\textbf{42.1\% (95\% CI: 34.3--50.2\%) are blends}: genuine eclipses
originating in a source other than the target, whose discrimination
requires spatial information absent from the light curve. This fraction is
a lower bound, since the one-pixel threshold (21~arcsec) does not resolve
closer neighbours (Sect.~\ref{sec:blends}).

\item Retraining with real false positives raises the AUC from 0.61 to
\textbf{$0.8006 \pm 0.0582$} under nested cross-validation. This value
is consistent with a heuristic bound derived from the blend fraction,
$\mathrm{AUC}_{\max} = 1 - f/2 \approx 0.79$, although the comparison is not
a strong test and the bound applies only to light-curve-only inputs of the
kind used here (Sect.~\ref{sec:ceiling}).

\item We document two manifestations of the same phenomenon within our own
work. An earlier version of the detector learned a geometric shortcut ---
the V shape of the primary transit --- introduced inadvertently by the
negative generator, in the sense described by \citet{Geirhos2020},
attaining a synthetic AUC of 0.99 while its real performance collapsed
(Sect.~\ref{sec:shortcut}). And a non-nested cross-validation protocol
inflated the AUC by \textbf{0.040}, a magnitude comparable to those
customarily invoked to establish superiority between architectures
(Sect.~\ref{sec:results-v4}).

\item Applied to 435 unresolved TOIs, the \textbf{CASCADE-USP} pipeline
yields 49 candidates scoring above 0.9, of which 29 are identified as
blends. Of the remaining 20, \textbf{one proves to be a validated planet
and 19 false positives}, documented individually by cause: chromatic
blends, spatial blends, eclipsing binaries on target, irreproducible
signals, astrometric binaries, and candidates lacking sufficient data.
\textbf{None of the 19 was distinguishable from a planet on the TESS light
curve alone} (Sect.~\ref{sec:application}).

\item The validated object ranks 26th of 435 by classifier score, while the
two highest-scoring candidates proved to be an eclipsing binary and a blend
with NFPP $= 1$. \textbf{The detector score does not rank by planetary
likelihood} among the survivors; its contribution is limited to
prioritisation (Sect.~\ref{sec:ordering}).

\item We introduce the Gaia RUWE as an additional vetting channel,
independent of photometry, centroid and contrast curve. Of the eighteen
surviving candidates with a published value, five (28\%) exceed the
threshold of 1.4, against the $\sim$12\% characteristic of the
\textit{Kepler} sample. Four of the five had already been rejected through
other channels --- astrometry confirming them independently --- and one was
flagged by astrometry alone, despite presenting the cleanest ground-based
record of the sample (Sect.~\ref{sec:taxonomy}).

\item We statistically validate \textbf{TOI-6662~b} (TOI-6662.01,
TIC 418343229) with FPP $= 0.0035$ and NFPP $= 0.00050$, satisfying the
criteria of \citet{Giacalone2021}, independently of the concurrent
validation by \citet{Collier2026}, with which it agrees in radius ratio,
scaled semi-major axis, planetary radius, classification and dependence on
the contrast curve. This agreement closes the CASCADE procedure with an
external check on a real planet. The transit fit yields
$\mathbf{R_p = 0.94 \pm 0.03\,R_\oplus}$ with $P = 0.2939$~d around an M
dwarf of $0.332\,R_\odot$. The planet receives 366 times the terrestrial
irradiation, reaching an equilibrium temperature of $\sim$1220~K, orbits at
1.19 times its tidal disruption limit --- which imposes a minimum density
consistent with a rocky composition --- and has an orbital decay timescale
of $\sim$20~$(Q'_\star/10^7)$~Gyr, stable against tidal inspiral
(Sect.~\ref{sec:planet}).

\item For detectors trained or evaluated on catalogued or simplistic
synthetic binaries, we recommend also reporting performance against real
TFOP false positives, as is already standard practice for the leading
classifiers (Sect.~\ref{sec:implications}).

\item Consistent with established practice, information absent from
single-band photometry --- colour, position, stellar field and astrometry
--- is essential for a substantial fraction of short-period false
positives, as reflected in recent practice with the inclusion of centroid data in ExoMiner
\citep{Valizadegan2022} and multiband photometry in TRICERATOPS+
\citep{GomezBarrientos2025} (Sect.~\ref{sec:limit}).

\end{enumerate}

\paragraph{Future work.}
Dynamical confirmation of TOI-6662~b requires a mass measurement; the
expected radial-velocity semi-amplitude, $\approx 1.6$~m~s$^{-1}$ for the
mass inferred from the mass--radius relation, is accessible to
high-stability near-infrared spectrographs given the brightness of the host
($T = 10.9$). The framework of CASCADE is independent of the period regime:
the centroid and dilution evaluation stages operate on any transit
candidate, so that adaptation to other regimes reduces to retraining the
classifier on the corresponding population.

\section*{Declarations}

\bmhead{Funding}
This research received no specific grant from any funding agency in the
public, commercial or not-for-profit sectors.

\bmhead{Competing interests}
The author declares no competing interests.

\bmhead{Ethics approval}
Not applicable.

\bmhead{Data availability}
The CASCADE code, the datasets used, and the intermediate results
supporting the figures and tables in this article are available from the
author upon reasonable request, and will be deposited in a permanent public
repository with the next revision of this work.

\bmhead{Software}
\texttt{lightkurve} \citep{Lightkurve2018},
\texttt{TRICERATOPS} \citep{Giacalone2021},
\texttt{juliet} \citep{Espinoza2019juliet},
\texttt{batman} \citep{Kreidberg2015},
\texttt{dynesty} \citep{Speagle2020},
\texttt{transitleastsquares},
\texttt{astroquery}, \texttt{astropy},
\texttt{numpy}, \texttt{scipy}, \texttt{scikit-learn},
\texttt{matplotlib}, \texttt{PyTorch}.

\bmhead{Use of AI tools}
The author used Claude (Anthropic), a large language model, as an
assistant for code development, literature search, and drafting and
language editing of the manuscript. The author designed and directed
the research, verified all analyses and references, and takes full
responsibility for the content of this article.

\bmhead{Acknowledgements}
This paper includes data collected by the TESS mission, obtained from the
Mikulski Archive for Space Telescopes (MAST). Funding for the TESS mission is
provided by NASA's Science Mission Directorate. This research has made use of
the Exoplanet Follow-up Observation Program (ExoFOP; DOI: 10.26134/ExoFOP5)
website, which is operated by the California Institute of Technology, under
contract with the National Aeronautics and Space Administration under the
Exoplanet Exploration Program. This work has made use of data from the European
Space Agency (ESA) mission \textit{Gaia} (https://www.cosmos.esa.int/gaia),
processed by the Gaia Data Processing and Analysis Consortium (DPAC). We
acknowledge the use of TESS Follow-up Observing Program (TFOP) data in this work.

\bibliography{references}

\end{document}